\documentclass[useAMS,usenatbib]{mn2e}

\usepackage{graphicx,subfigure,color}
\usepackage[total={17.8cm,24.0cm},centering]{geometry}
\usepackage{times}

\def\hi{H{\small I}}
\def\sauron{{\tt SAURON}}

\def\atlas{{{ATLAS}}$^{\rm 3D}$}
\def\kms{km s$^{-1}$}
\def\co{CO}
\def\tirific{Ti{R}iFiC}
\def\vmax{V$_{\rm flat}$}
\def\arcsec{$^{\prime \prime}$}
\definecolor{Mygrey}{gray}{0.75}

\newcommand{\ltsimeq}{\raisebox{-0.6ex}{$\,\stackrel{\raisebox{-.2ex}{$\textstyle <$}}{\sim}\,$}}

\usepackage[compact]{titlesec}
\titlespacing{\section}{0pt}{*2}{*1}

\title[The CO Tully-Fisher relation of early-type galaxies]{The \atlas\ project -- V. The CO Tully-Fisher relation of early-type galaxies}
\author[Timothy A. Davis et al.]{Timothy A. Davis$^1$\thanks{Email: timothy.davis@astro.ox.ac.uk}, Martin Bureau$^1$, Lisa M. Young$^2$\thanks{Adjunct Astronomer with NRAO}, Katherine Alatalo$^3$, Leo Blitz$^3$,
 \newauthor 
 Michele Cappellari$^1$, 
 Nicholas Scott$^1$,
 Maxime Bois$^{4,5}$, 
 Fr\'ed\'eric Bournaud$^{6}$, 
 \newauthor 
 Roger L. Davies$^1$, 
 P. Tim de Zeeuw$^{4,7}$, 
 Eric Emsellem$^{4,5}$, 
 Sadegh Khochfar$^8$, 
 \newauthor 
 Davor Krajnovi\'c$^4$, 
 Harald Kuntschner$^{9}$, 
 Pierre-Yves Lablanche$^{5}$, 
   \newauthor 
 Richard M.\,McDermid$^{10}$, 
   Raffaella Morganti$^{11,12}$, 
  Thorsten Naab$^{13}$, 
Tom Oosterloo$^{11,12}$, 
  \newauthor 
    Marc Sarzi$^{14}$, 
  Paolo Serra$^{11}$, 
  and Anne-Marie Weijmans$^{15}$\thanks{Dunlap Fellow}\\
$^1$Sub-department of Astrophysics, Department of Physics, University of Oxford, Denys Wilkinson Building, Keble Road, Oxford, OX1 3RH, UK\\
$^2$Physics Department, New Mexico Institute of Mining and Technology, Socorro, NM 87801, USA\\
$^3$Department of Astronomy, Campbell Hall, University of California, Berkeley, CA 94720, USA\\
$^4$European Southern Observatory, Karl-Schwarzschild-Str. 2, 85748 Garching, Germany\\
$^5$Universit\'e Lyon 1, Observatoire de Lyon, Centre de Recherche Astrophysique de Lyon and Ecole Nationale Sup\'erieure de Lyon,\\\,\, \,9 avenue Charles Andr\'e, F-69230 Saint-Genis Laval, France\\
$^6$Laboratoire AIM Paris-Saclay, CEA/IRFU/SAp Ð CNRS Ð Universit\'e Paris Diderot, 91191 Gif-sur-Yvette Cedex, France\\
$^7$Sterrewacht Leiden, Leiden University, Postbus 9513, 2300 RA Leiden, the Netherlands\\
$^8$Max Planck Institut f\"ur extraterrestrische Physik, PO Box 1312, D-85478 Garching, Germany\\
$^{9}$Space Telescope European Coordinating Facility, European Southern Observatory, Karl-Schwarzschild-Str. 2, 85748 Garching, Germany\\
$^{10}$Gemini Observatory, Northern Operations Centre, 670 N. A`ohoku Place, Hilo, HI 96720, USA\\
$^{11}$Netherlands Institute for Radio Astronomy (ASTRON), Postbus 2, 7990 AA Dwingeloo, The Netherlands\\
$^{12}$Kapteyn Astronomical Institute, University of Groningen, Postbus 800, 9700 AV Groningen, The Netherlands\\
$^{13}$Max-Planck-Institut f\"ur Astrophysik, Karl-Schwarzschild-Str. 1, 85741 Garching, Germany\\ 
$^{14}$Centre for Astrophysics Research, University of Hertfordshire, Hatfield, Herts AL1 9AB, UK\\
$^{15}$Dunlap Institute for Astronomy \& Astrophysics, University of Toronto, 50 St. George Street, Toronto, ON M5S 3H4, Canada 
}
\begin{document}
\date{Accepted 2010 January 4. Received 2010 December 27; in original form 2010 Novomber 02 }
\pagerange{\pageref{firstpage}--\pageref{lastpage}} \pubyear{2009}
\maketitle
\label{firstpage}
\begin{abstract}

We demonstrate here using both 
single-dish and interferometric observations that CO molecules are an excellent kinematic tracer, even in high-mass galaxies, allowing us to investigate for the first time
the CO Tully-Fisher relation of early-type galaxies. 
We compare the Tully-Fisher relations produced using both single-dish and interferometric data and various inclination estimation methods, and evaluate the use of the velocity profile shape as a criterion for selecting galaxies in which the molecular gas extends beyond the peak of the rotation curve. 
We show that the gradient and zero-point of the best-fit relations are robust, independent of the velocity measure and inclination used, and agree with those of relations derived using stellar kinematics.
We also show that the early-type CO Tully-Fisher (CO-TF) relation is offset from the CO-TF of spirals by 0.98 $\pm$ 0.22 magnitude at $K_{\rm s}$-band, in line with other results.
The intrinsic scatter of the relation is found to be $\approx$0.4 magnitudes, similar to the level found in the spiral galaxy population.
Next generation facilities such as the Large Millimeter Telescope (LMT) and the Atacama Large Millimeter/Sub-millimeter Array (ALMA) should allow this technique to be used in higher-redshift systems, providing a simple new tool to trace the mass-to-light ratio evolution of the most massive galaxies over cosmic time.
\end{abstract}

\clearpage
\section{Introduction}
The Tully-Fisher relation \citep[TFR;][]{Tully:1977p2161} of spiral galaxies has proved itself 
to be one of the most important correlations in extragalactic astrophysics. For example, its use as a distance measure is vital in extending the cosmic distance ladder, allowing the scale of structures in the nearby universe to be determined and studied. The underlying cause of this relation between luminosity and 
rotational velocity is usually interpreted as the product of a relatively constant total (luminous plus dark) mass-to-light ratio (M/L) in the local spiral galaxy population \citep{Gavazzi:1993p2448, Zwaan:1995p2450}, and hence a strong coupling between dark and luminous mass. Studying the slope and zero-point of the TFR is thus also a powerful probe of the M/L evolution of galaxies \citep[e.g.][]{Phillipps:1989p3235,Sprayberry:1995p3236,Bell:2001p3237}.

The TFR as introduced by \cite{Tully:1977p2161} uses rotation velocities derived from \hi\ line-widths in spiral galaxies, which should approximate the true projected circular velocity of the galaxies as long as the \hi\ distributions are relaxed and reach into the flat parts of the galaxy rotation curves. Over the years, however, the TFR has gradually been recognised as denoting the empirical relationship between the luminosity and rotation velocity of galaxies (generally disc galaxies) as measured with a variety of kinematic tracers, using differing techniques at various wavelengths. 

Studying the TFR in lenticular and elliptical galaxies (collectively referred to as early-type galaxies; ETGs) is problematic, as they do not all possess extended relaxed atomic gas distributions. Indeed, it has been shown that even where \hi\ is present in early-type galaxies it can be disturbed, and hence the measured line-widths are in some cases unrelated to galaxy properties \citep[e.g.][]{Morganti:2006p1934,Williams2010}. Although widespread, the ionised gas is generally faint, with significant pressure support; and is thus not ideal either \citep{Bertola:1995p3239}. The TFR of early-type galaxies is nevertheless important, as these galaxies are believed to have turbulent formation histories and the TFR may give clues about their assembly and evolution. For example, many authors have suggested that S0 galaxies have avoided violent interactions, and are the faded descendants of high-redshift spirals \cite[e.g.][]{Dressler:1980p2456,Dressler:1997p2485}. In this scenario, S0 galaxies become dimmer whilst keeping the same dynamical mass, leading to an offset TFR.

Due to the problems listed above, stellar tracers of galactic rotation are generally used in ETGs. Due to the importance of pressure support in these systems, however, stellar dynamical modeling or an asymmetric drift correction is required to extract the true circular velocities. This adds additional systematic uncertainties to the already challenging stellar kinematic observations (which must reach sufficiently large radii). Recent analyses of ETGs have suggested that S0 galaxies do indeed have a measurable offset from the spiral TFR, of around 0.5 - 1.0 mag at $K$-band \citep{Neistein:1999p3044,Bedregal:2006p2087,Williams2010}. \cite{Magorrian:2001p3045}, \cite{Gerhard:2001p3046} and \cite{DeRijcke:2007p2492} have all considered extending this approach to construct TFRs that include elliptical galaxies, and find that these are also offset from the spiral TFR, by 0.5 to 1.5 mag at optical wavelengths.

In this work we consider the use of carbon monoxide (CO) as a tracer of the circular velocity, of fast-rotating early-type galaxies. CO is thought to be free of many of the problems that beset \hi\ in early-types. 
As part of the \atlas\ survey of ETGs \citep[][hereafter Paper I]{Cap2010} we have found that molecular gas is reasonably abundant in early-type galaxies \citep[][hereafter Paper IV]{Young2010}
, with $\approx$22\% of early-type galaxies in the local volume ($<$42 Mpc distant) containing a substantial molecular gas reservoir. The detection rate of molecular gas is also independent of galaxy luminosity and mass (the two most important TFR parameters), providing a direct and unbiased probe of the potential in high-mass galaxies, which are often \hi\ poor (Serra et al., in preparation). The molecular gas is likely to be relaxed in most galaxies under study, even in clusters where \hi\ may be undergoing ram-pressure stripping \citep{Toribio:2009p2676}, due to the short dynamical timescales in the central parts of the galaxies, where it is usually found \citep[e.g. ][]{Wrobel:1992p982,Young:2002p943,Young:2008p788,Crocker:2008p946,Crocker:2009p3262,Crocker2010}. Therefore CO provides a powerful, directly observable measure of the circular velocity of galaxies of all masses and morphological types, irrespective of enviroment. Additionally, the small beamsizes of the mm-wave single-dish telescopes used to detect CO eliminate source confusion, at least in nearby galaxies, while current interferometers routinely yield arcsecond angular resolutions. 

The possibility of using CO line-widths to investigate the Tully-Fisher relation was first explored by \cite{Dickey:1992p2418}, and this method has since been used by various authors to investigate the CO TFR of spiral and irregular galaxies \citep[e.g.][]{Schoniger:1994p2416,Schoeniger:1997p2424,Tutui:1997p2423,Lavezzi:1998p2411,Tutui:2001p3401} and quasars \citep{Ho:2007p2426}. In spiral galaxies, the CO velocity widths obtained are directly comparable to those found in \hi\ \citep{Lavezzi:1997p2421}.  \cite{Young:2008p788} have also demonstrated that CO is a good tracer of the circular velocity in a small sample of ETGs,  from a comparison with detailed dynamical modeling of the stellar kinematics, but this has yet to be demonstrated for a larger sample.

One possible complication introduced when using CO as a dynamical tracer is that the molecular gas in early-types is usually confined to the inner regions. This means that we are only able to probe the kinematics where dark matter does not yet play a significant role. This is the opposite of \hi\ TF analyses, which are generally probing \hi\ at large radii, in the dark matter-dominated part of the rotation curve. As such, a TFR measured in the central region could be considered more similar to the Faber-Jackson (central velocity dispersion - luminosity) relation \citep{Faber:1976p3242}. The ETG CO TFR will therefore only be directly comparable to TF results from \hi\ if the `disc-halo conspiracy' , where dark matter flattens the rotation curve at a similar velocity to that found (from luminous material) in the inner regions, also holds for early-type galaxies \citep{Kent:1987p2832,Sancisi:2004p2830,Gavazzi:2007p2831}.

Some authors have suggested that 
there is a change in the slope of the TFR for high-mass disc galaxies, brighter than an absolute $K$-band magnitude of $\approx$-23.75 \citep{Peletier:1993p3082,Verheijen:2001p3081,Noordermeer:2007p3078}. A break in the TFR at a similar position is also found for elliptical galaxies \citep{Gerhard:2001p3046,DeRijcke:2007p2492}  It has been suggested that this break occurs because many massive galaxies have declining circular velocity profiles, whereas low mass galaxies have relatively flat circular velocity curves. In other words, the aforementioned disc-halo conspiracy is not perfect in ETGs, and many have a local rotation velocity peak at small radii. The radius where one measures the velocity hence becomes important \citep[see][]{Noordermeer:2007p3078}. Measures of maximal rotation produce a much larger break than measures of the asymptotic rotation velocity, supporting this interpretation. Clearly, the existence of a break in the TFR could lead to systematic biases when deriving distances or probing galaxy evolution. Fortunately, however, we show in this paper that our CO velocities are consistent with measures of the rotation beyond the peak of the galaxy circular velocity curves. 

This work, to the best of the author's knowledge, represents the first attempt to create a CO TFR for early-type galaxies. One can identify two major pitfalls that need to be overcome for this approach to be successful. Firstly, as discussed above, it has been shown that the CO in ETGs is often very centrally concentrated, and hence in some galaxies it may not reach beyond the peak of the rotation curve. Identifying such cases is critical to obtain a useful TFR. Secondly, it has been shown that the molecular gas in early-type galaxies is often misaligned with respect to the stars (e.g. \citealt{Young:2002p943,Schinnerer:2002p981,Young:2008p788,Crocker:2009p3262}a; \citealt{Crocker:2008p946}b). In these cases the inclination of the stars is not useful, and we require an estimate of the inclination of the molecular gas itself in order to de-project its observed rotation velocity.     

Our goal in this paper is therefore to explore and demonstrate that, despite the potential pitfalls highlighted above, ETGs do appear to follow a robust luminosity-rotational velocity relation, consistent with that measured using other tracers at the same radii. Although technical aspects will certainly be improved and the interpretation of this relation remains uncertain, an ETG CO TFR is a tantalising and promising tool for galaxy evolution studies, worthy of further consideration. In Section 2 of this paper we discuss the data used in this work, outlining the observations in Sections 2.3 and 2.4, the methods for estimating the inclination in Section 2.5, and our method for extracting the circular velocity from observed CO line-widths in Section 2.6. In Section 3 we compare these velocities with other measures such as circular velocities derived from dynamical models, and show that it is possible to use simple criteria to select galaxies where CO is a good tracer of the circular velocity beyond the peak of the galaxy rotation curve. In Section 4 we present our ETG CO TFRs, and explore how different data and inclination estimates affect the resulting relations. In Section 5 we more fully discuss our results and compare them with previous TFR results derived using other tracers. We summarise our conclusions in Section 6.

\section{Data}
\subsection{Sample}
The sample used here is composed of the galaxies detected in \co(1-0) emission in Paper IV. \co(1-0) and (2-1) were observed in all galaxies from the complete, volume-limited \atlas\ sample of 260 lenticular and elliptical galaxies (see Paper I for full details).
Upper limits for non-detections are typically in the range $10^7$-$10^8$ $M_{\odot}$, depending on the distance, assuming a standard CO-to-H$_2$ conversion factor. The detected galaxies used here range in molecular gas masses from $10^7$ to $10^9$ $M_{\odot}$, with molecular gas mass fractions between $10^{-4}$ and $10^{-1}$ $M_{\odot}$/$L_{K\rm{s}}$, where $L_{K\rm{s}}$ is the $K_{\rm{s}}$-band luminosity from the Two Micron All Sky Survey (2MASS) \citep{Jarrett:2000p2407,Skrutskie:2006p2829}. Only fast-rotating galaxies were detected.

Of the 52 detections by Paper IV, 4 are only detected in \co(2-1), 
and a further 8 had an insufficient signal to noise ratio to perform the analysis required here, leaving 40. In addition to these we include the ETGs IC2099, NGC\,4292 and NGC\,4309, which were detected in \co\ as part of the initial \atlas\ survey, but were later removed from the sample because they are too faint at $K_{\rm{s}}$-band to meet the final sample selection criteria. This should not affect the TFR. This leaves a total of 43 galaxies, the properties of which are listed in Table \ref{table}. 

\subsection{Photometric data}
The $K$-band magnitudes used in this paper are $K_{\rm s,total}$ from 2MASS \citep{Jarrett:2000p2407,Skrutskie:2006p2829}.  
These $K_{\rm s,total}$ magnitudes are measured over large apertures, to include the total flux from the galaxy using the techniques developed in \cite{Kron:1980p3008} and curves-of-growth (see \citealt{Jarrett:2000p2407} for further details). They have been widely used in the astronomical community and are found to be robust to $\approx$0.1 mag \citep{Noordermeer:2007p3078}.

\subsection{CO single-dish data}
\label{singledish}
As reported in Paper IV, the Institut de Radioastronomie Millim\'etrique (IRAM) 30m telescope at Pico Veleta, Spain, was used for simultaneous observations of
\co(1-0) and \co(2-1) in our galaxies. The primary beam FWHM is 23\arcsec\ and 12\arcsec\ for \co(1-0) and \co(2-1) respectively.  
The filterbank back-end gave an effective total bandwidth of 512 MHz ($\approx$1330 \kms) and a raw spectral
resolution of 1 MHz (2.6 \kms) for \co(1-0). The system temperatures ranged from 190 to 420 K for \co(1-0). The time on source was interactively adjusted
so that the final, co-added \co(1-0) spectrum for each galaxy had a rms noise level of  $\approx$3.0 mK T$_{\rm a}^{*}$ ($\approx$19 mJy) per binned 31 \kms\ channel. 
For further details, see Paper IV.

The \co(1-0) and \co(2-1) spectra for each detected galaxy, binned to 31 \kms\ channels, were analyzed to find the velocities at which the flux drops to 20\% of the peak height. Each spectrum was read into a routine which, working outwards from the galaxy systemic velocity, locates the first channel where the velocity drops to 20\% of the peak value. The measurements were inspected by eye to ensure noise peaks and troughs were not affecting the result. This is analogous to the method used by \cite{Tully:1977p2161} for measuring \hi\ line-widths. The resulting velocity width is henceforth denoted W$_{20}$. If the molecular gas is a good tracer of the circular velocity in our galaxies, then the line-width should approximate twice the projected rotational velocity, if the molecular gas distribution reaches beyond the peak of the galaxy rotation curve \citep{Dickey:1992p2418}.

\begin{figure}
\begin{center}
\includegraphics[scale=0.525,angle=0,clip,trim=1cm 0cm 0.5cm 1cm]{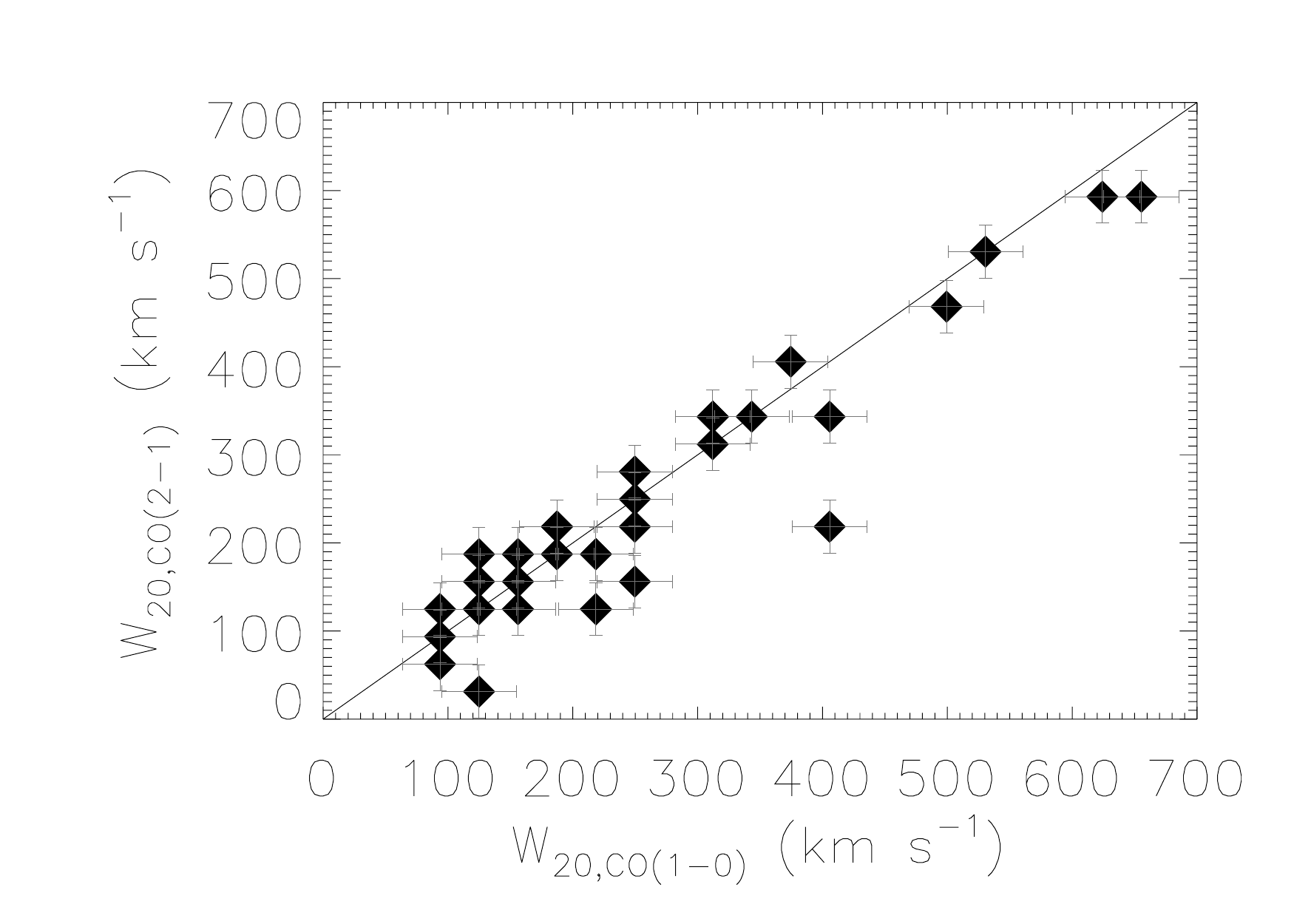}
\end{center}
\caption{\small A comparison between \co(1-0) and \co(2-1) single-dish line-widths, measured as described in Section \ref{singledish}. The solid line shows the one-to-one relation. Linewidths are discretized due to the 30 \kms\ channel width.}
\label{co10v21}
\end{figure}

Figure \ref{co10v21} shows a comparison between W$_{20}$ line-widths measured from \co(1-0) and \co(2-1) spectra. One might expect it to be easier to measure line-widths using \co(2-1), as Paper IV have shown that this transition is usually 1--4 times brighter than \co(1-0) (due to beam dilution and/or intrinstic effects), but Figure \ref{co10v21} shows that the agreement between the two measurements is generally good. However, almost all the scatter is below the 1:1 line. In these cases, the molecular gas likely extends further than the $\approx$12\arcsec\ FWHM of the \co(2-1) beam, and the CO(2-1) line-width measurements are systematically biased low. In Section \ref{tiltringsec}, we in fact show that in some cases even the $\approx$23\arcsec\ CO(1-0) beam is not extended enough to retrieve the full velocity width. The number of galaxies for which this is a concern is small, however, and in the rest of this paper we will therefore use line-widths measured from the \co(1-0) spectra.

 The instrumental dispersion is small compared to our 31 \kms\ channels, so we did not correct the line-widths for it. 
We also chose not to use corrections for turbulence or line-broadening, as these are generally derived by comparison with \hi\ line-widths in spirals \citep{Lavezzi:1998p2411,Tully:1985p2513,Tutui:1999p2518} and hence may not be applicable to early-type galaxies. The velocity dispersion in the gas is expected to be small \citep[e.g.][]{Okuda:2005p3259}. 
In Section \ref{modelcurves} we attempt to quantify any biases introduced by ignoring these corrections. The derived W$_{20}$ velocity widths are estimated to be robust to within half a channel width, $\approx$15 \kms, and are listed in Table \ref{table}.

\subsection{Interferometric data}
\label{intervflat}

As part of the \atlas\ survey, all \co(1-0) detections with an integrated flux greater than about 19 Jy \kms\ that do not have interferometric maps available in the literature are to be observed with the Combined Array for Research in mm-wave Astronomy \citep[CARMA;][]{Bock:2006p2806}. Full details of this ongoing interferometric survey can be found in Alatalo et al. (in preparation), but we summarize the observations here.  

Observations of the sample galaxies have been ongoing since early 2008, mainly in the D-array configuration, providing a spatial resolution of 4-5\arcsec.
\co(1-0) has so far been observed using narrow-band correlator configurations, providing at least 3 raw channels per 10 \kms\ binned channel whilst ensuring adequate velocity coverage for all galaxies.
Bright quasars were used to calibrate the antenna-based gains and for passband calibration. The data were calibrated and imaged using the `Multichannel Image Reconstruction, Image Analysis and Display' (MIRIAD) software package \citep{Sault:1995p2768}. Total fluxes were compared with the IRAM 30m single-dish observations to ensure that large proportions of the fluxes were not being resolved out. 

A total of 22 galaxies included in this work have been observed with CARMA so far (IC\,0676, IC\,719, IC\,1024, NGC\,1222, NGC\,1266, NGC\,2764, NGC\,2824, NGC\,3626, NGC\,3665, NGC\,4119, NGC\,4292, NGC\,4324, NGC\,4429, NGC\,4435, NGC\,4694, NGC\,4710, NGC\,5379, NGC\,6014, NGC\,7465, PGC\,058114, UGC\,06176 and UGC\,09519). We also include here the galaxies for which data is already available from the literature, mostly from SAURON survey \citep{deZeeuw:2002p1496} follow-ups. These are  NGC\,3032, NGC\,4150, NGC\,4459 and NGC\,4526 \citep{Young:2008p788}; NGC\,2685 \citep{Schinnerer:2002p981}; NGC\,2768 \citep{Crocker:2008p946}; NGC\,0524, NGC\,3489, and NGC\,4477 \citep{Crocker2010}. This makes for a total of 31 sample galaxies that have interferometric data.

The data cubes were summed spatially 
to measure total fluxes and revised values for W$_{20}$, serving as consistency checks on the values derived from the single-dish data.
The large primary beam of the arrays also allow us to make better measurements for the small subset of galaxies in which a substantial part of the \co\ distribution was missed by the 23\arcsec\ beam of the IRAM 30m telescope, such as in IC676, NGC\,4324, NGC\,4477, NGC\,4710, NGC\,7465 and PGC\,058114. This can be due to pointing errors or molecular gas distributions that extend beyond the beam. The integration time for the single-dish CO observations was adjusted to obtain a fixed noise level- and as such galaxies which are only detected at $\approx$5$\sigma$ will have rather uncertain W$_{20}$ line-widths. The interferometric observations in these cases (NGC\,2685, NGC\,2768, NGC\,3489, NGC\,4477) will provide a better constraint on the line-widths.   
 The revised W$_{20}$ values are listed in Table \ref{table}, but for consistency these velocities are only used in the TFRs presented in Sections \ref{tiltringsec} and \ref{best}.

\subsection{Inclination correction}
\label{inccor}
The measured quantity W$_{20}$ is a projection of the gas velocity into the line-of-sight. If this is to be used for a TF analysis it must be deprojected.
Many methods for inclination measurement are available. Starting with the simplest, we compare various method below in order of increasing complexity, to allow future CO TF surveys to select the optimum method for their needs.
 The methods presented in Sections \ref{galaxial} and \ref{galdust} use only single-dish molecular data, whilst the method presented in Section \ref{inter} requires interferometric maps.

\subsubsection{Galaxy axial ratio}
\label{galaxial}
A rough measurement of inclination can be obtained by estimating the axial ratio of the stellar distribution of the host galaxy from imaging data: 
\begin{equation}
\label{comaxial}
i_{\rm b/a} = \cos^{-1}\left(\sqrt{\frac{q^2-q_0^2}{1-q_0^2}}\right),
\end{equation}
\noindent where $q$ is the ratio of the semi-minor ($b$) to the semi-major ($a$) axis of the galaxy, and $q_0$ is the intrinsic axial ratio when the galaxy is seen edge-on ($q_0 \equiv c/a$). $q_0$ is often assumed to be 0.2 in disc galaxies \citep{Tully:1977p2161}, but early-types can have large bulge-to-disc ratios leading to a large uncertainty in any assumed value of $q_0$. 

Various lines of enquiry suggest a mean $q_0$ value of $\approx$0.34 for the fast-rotators in the \atlas\ sample (Weijmans et al., in preparation), which we adopt here. Intrinsic scatter around this value will introduce an artificial increase in the TFR scatter, but the effect is very small. Indeed, we stress that this $q_0$-related inclination correction is only significant in highly inclined galaxies, where fortunately inclination errors have the smallest effect on the de-projected velocities and thus the derived TFR. 
For example, assuming $q_0=0$ for the current sample would only change the zero-point of the derived TFR by 0.02 dex along the velocity axis, and it does not change the gradient. 

It is important to note that using this method to correct the CO velocity widths for a TF analysis implicitly assumes that the CO is distributed in the same plane as the galaxy major-axis. Recent work by Davis et al., (in preparation) however suggests that a sizable proportion of early-type galaxies have molecular gas misaligned with respect to the stellar kinematic axis, so in these cases one expects this method to introduce additional artificial scatter in the TFR.

The value of $i_{\rm b/a}$ obtained by assuming $q_0$=0.34 for each galaxy is listed in Table \ref{table}. These values were calculated using Equation \ref{comaxial} with the mean galaxy axial ratio reported in the NASA/IPAC Extragalactic Database (NED). These values are the mean of the measurements from the Sloan Digital Sky Survey (SDSS) in the $r$ band, 2MASS at $K_s$ band \citep{Skrutskie:2006p2829}, and blue and red filters from both the Third Reference Catalogue of Bright Galaxies \citep[RC3;][]{deVaucouleurs:1991p2406}, and the Uppsala General Catalogue of galaxies \citep[UGC/POSS;][]{Nilson:1973p3009} where available.  The exceptions to this are NGC\,3665, where the axial ratio is based only on the superior SDSS imaging, and PGC\,058114, where the mean value of the axial ratio from the HyperLeda database \citep{Paturel:2003p3431} has been used, due to a larger number of available measurements. When these inclinations are compared to those estimated from the axial ratios of \cite{Kraj2010} (Paper II; derived by calculating the moments of inertia of the surface brightness distribution from the SDSS and INT r-band images) for the galaxies in the \atlas\ sample they are found to agree well.

\subsubsection{Dust axial ratio}
\label{galdust}
It should be possible to obtain more accurate inclination estimates from fitting ellipses to dust highlighted in unsharp-masked optical images of the galaxies. This method has several advantages over galaxy axial ratios. Dust distributions typically have very small vertical scale-heights, hence Equation \ref{comaxial} with $q_0$=0 should yield a good estimate of their inclinations. It has also been shown that CO and dust in early-type galaxies are usually spatially coincident \citep{Young:2008p788,Crocker:2008p946,Crocker:2009p3262,Crocker2010}, hence the inclination of the dust ($i_{\rm dust}$) should trace the true inclination of the molecular gas. This is especially useful where the molecular gas and stars are misaligned.

\begin{figure}
\begin{center}
\subfigure[NGC\,2824]{\includegraphics[scale=0.4,angle=0,clip,trim=8.5cm 4.5cm 8.5cm 3cm]{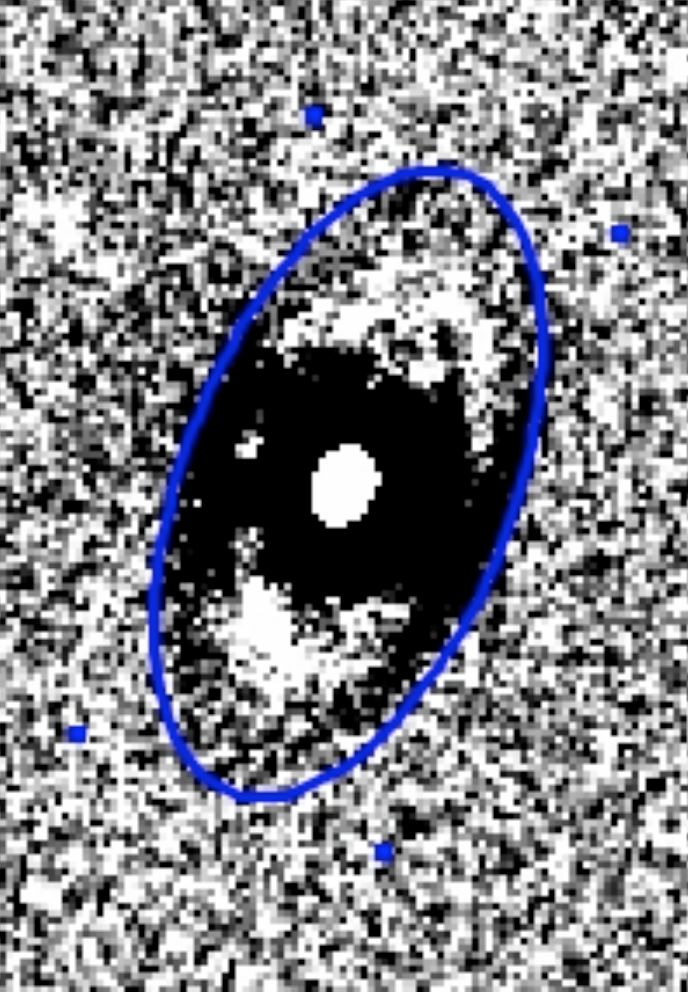}}
\subfigure[NGC\,4526]{\includegraphics[scale=0.4,angle=0,clip,trim=8.5cm 4.5cm 8.5cm 3cm]{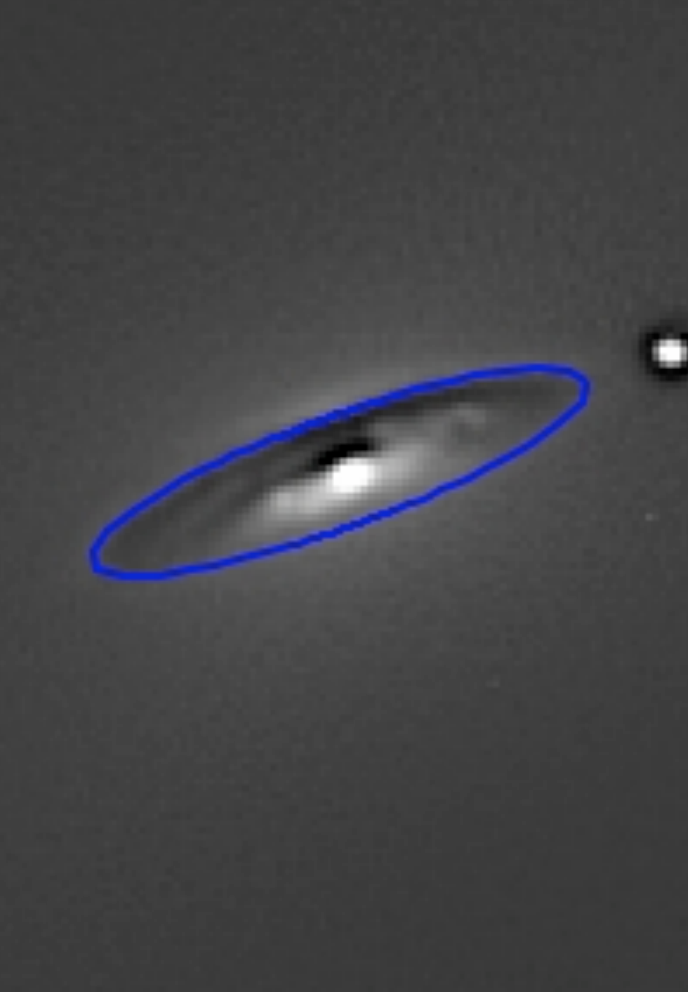}}
\subfigure[NGC\,3626]{\includegraphics[scale=0.4,angle=0,clip,trim=8.5cm 4.5cm 8.5cm 3cm]{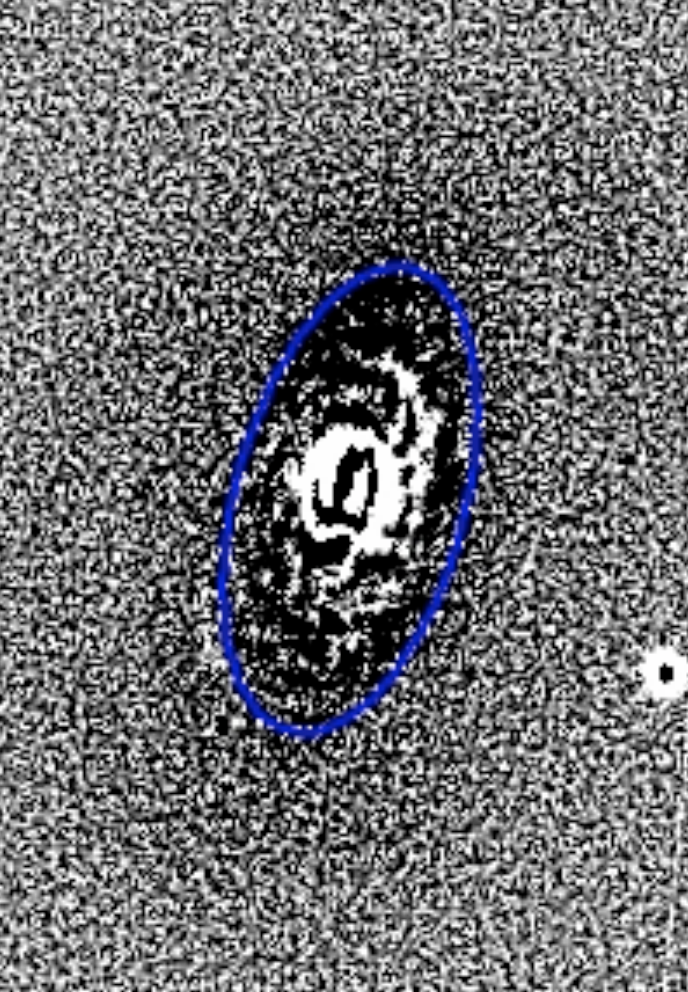}}
\subfigure[NGC\,5379]{\includegraphics[scale=0.4,angle=0,clip,trim=8.5cm 4.5cm 8.5cm 3cm]{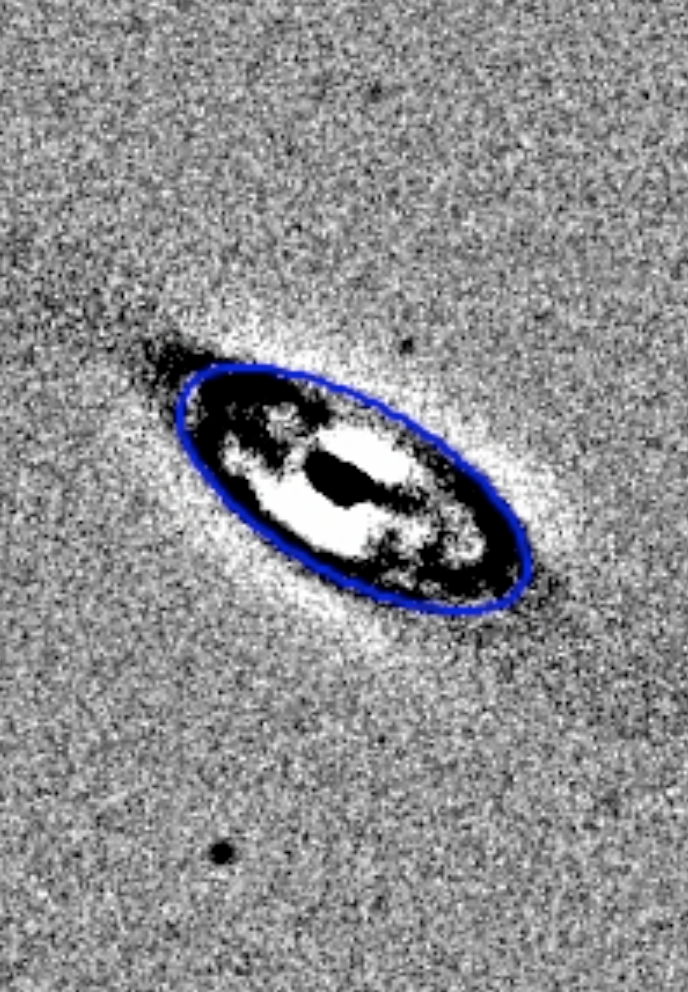}}
\end{center}
\caption{\small Four examples of unsharp-masked images picking up dusty spiral structures and rings. Overlaid in blue are the best fit ellipse from which we estimate the dust inclination.}
\label{dustmaps}
\end{figure}

Unsharp-masked dust maps were created for all the galaxies in this work, preferentially from archival Hubble Space Telescope (HST) images, or where these were not available from SDSS $g$-band \citep{AdelmanMcCarthy:2008p2834} or \atlas\ Isaac Newton Telescope Wide Field Camera (INT-WFC) $r$-band images (Scott et al., in preparation). Ellipses were then fitted by eye to the resulting maps and the inclination of the dust calculated from Equation \ref{comaxial} with $q_0$=0. Four example galaxies are shown in Figure \ref{dustmaps}.

One of the galaxies (NGC\,4753) is well studied in the literature, showing many thin filamentary dust features that are well fitted by a model featuring an inclined disc, twisted by differential precession, with an average $i_{\rm dust}$=75 $\pm\,3^{\circ}$ \citep{SteimanCameron:1992p2507}. We adopt this value here. A full list of $i_{\rm dust}$ values can be found in Table \ref{table}.

\subsubsection{Molecular gas modeling}
\label{inter}

Tilted-ring analyses have been used to determine the geometric and kinematic parameters of neutral and molecular gas discs in interferometric data for many years \citep[e.g. ][]{Rogstad:1974p2774,Christodoulou:1988p3010,Koribalski:1993p3012}. In this work, we have made use of the Groningen Image Processing System (GIPSY) add-on package \tirific, described in \cite{Jozsa:2007p2673}, to fit tilted-ring models to observed data cubes. Due to the small number of independent synthesized beams across our typical sources, an unwarped disc model was fitted where the inclination, position angle, systemic velocity and kinematic centre of the gas were varied globally, and only the velocity and surface brightness were allowed to vary radially (i.e. for each ring). 
Full details of the fits are reported in Davis et al., (in preparation) but the inclination measures ($i_{\rm mol}$) used in this work are listed in Table \ref{table}.
The error quoted for each galaxy is the maximum inclination difference found if each side of the velocity distribution is supplied as an input to the fitting program separately. 

We have attempted the analysis described above for all the mapped galaxies in our sample where with have adequate spatial resolution, with one exception where we felt the literature value was sufficient. NGC\,2685 has dust lanes which look polar on the north-east side, and are coincident with \hi, H$\alpha$ and CO emission \citep{Schinnerer:2002p981}. This system has been revealed to feature a coherent, extremely warped disk \citep{Jozsa:2009p3232}, for which we adopt the published average inclination value of $i_{\rm mol}$=69 $\pm\,9^{\circ}$.

\subsection{W$_{20}$ linewidths}

\label{velwidths}

The velocity widths 
 calculated as described in Sections \ref{singledish} and \ref{intervflat} are an approximation to (twice) the maximum rotation velocity of the \co. To be useful, 
these \co\ measurements must reach beyond the peak of the galaxy rotation curves. \cite{Young:2008p788} suggest that molecular gas in early-types is often centrally concentrated and hence this requirement is not trivially fulfilled.

One sign of a tracer having reached a flat part of a galaxy rotation curve is a classic `double-horned' velocity profile. \cite{Lavezzi:1997p2868} discuss in detail the problem of recovering velocity widths from CO observations, given the variety of profile shapes typically seen in millimeter observations.
They show that CO profile shapes are affected by the extent of the emitting gas, the beam size of the telescope, pointing errors and the optical depth of the molecular material. CO line-widths are shown to accurately retrieve \vmax\ in the majority of cases where the flaring parameter R$_{2,5}$ (the ratio of the velocity widths measured at 20 and 50\% of the peak) is less than 1.2 \citep{Lavezzi:1997p2868,Lavezzi:1998p2411}. This criterion effectively selects galaxies that show a boxcar, or double-horned, profile shape with sharp edges and rejects those with a more Gaussian profile. We follow this methodology in this work, exclusively using galaxies with double-horned/boxcar/sharp edged profiles, which we call 'boxy' from here on in, assuming that in these systems the measured gas velocity width will approximate the velocity beyond the peak of the circular velocity curve.

The limited signal-to-noise ratio of the spectra we possess, however, means that the value of the flaring parameter discussed above is often rather uncertain. 
We performed several other tests, attempting to find a good automatic way of classifying boxy velocity profiles. These included fitting single gaussians and selecting those galaxies that shown significant structure in the residuals, and fitting Gauss-Hermite polynomials and looking for a significant kurtosis (h$_4 <$ 0). Both of these methods worked well for wide lines and high signal to noise cases, but in the case of narrow lines, or low signal to noise the fits were not well constrained. 
In these cases the different methods often produced contradictory results, and/or did not agree with a careful classification by eye. In future surveys with better data quality, or large numbers of objects (so one can afford to remove low signal to noise detections) these methods will likely be highly useful, but in this current work, with a limited number of objects we prefer a careful classification by eye.

Even by eye, some cases are hard to classify, especially when the profiles are narrow as the galaxies are fairly face on, or if the signal-to-noise ratio is low. We cautiously use only those galaxies with clear boxy profiles here, flagging the 12 galaxies with uncertain profiles shapes. 

Using this profile shape criterion is likely to introduce some uncertainties, for example galaxies with a molecular ring will have double-horned profiles without the CO necessarily having reached beyond the peak of the rotation curve. Similarly, galaxies that have strong resonances may have their gas distribution truncated at the turnover radius, causing them to not display a classic boxy shape even though the edge of the distribution is rotating at the turnover velocity.
In Section \ref{modelcurves} we use the dynamical models available to us from the \atlas\ survey to quantify these biases.

\subsubsection{Combining interferometric and single-dish data}
\label{bestinc}
Finally, we can combine our single-dish and interferometric observations, using all available data to increase the number of galaxies  with high-quality measurements available. When available, we thus always use interferometric velocity widths and, when well constrained, interferometric inclinations (quantitatively, we select galaxies with inclination errors less than 6$^\circ$). For the remaining galaxies we preferentially use dust inclinations, or failing this, $b$/$a$ inclinations. We can also use the interferometric data when available to help ascertain which galaxies have reached beyond the peak of the rotation curve, in conjunction with the usual single-dish boxy criterion. This hybrid measure represents our best estimate of the inclination and velocity width. The adopted best inclinations are listed in Table \ref{table}.

\begin{table*}
\caption{Parameters of the \atlas\ early-type CO-rich galaxies used in this paper. }
\begin{tabular*}{1\textwidth}{@{\extracolsep{\fill}}r r r c c c c c c c c c c}
\hline
Galaxy & $T$-type & W$_{\rm 20,SD}$ & W$_{\rm 20,inter}$ & $i_{\rm b/a}$ & $i_{\rm dust}$ & $i_{\rm tilted-ring}$ & $i_{\rm best}$ & M$_{K\rm{s}}$ & Dhorn & Dhorn & Tel. & Ref. \\
 & & (\kms) & (\kms) & (deg) & (deg) & (deg) & (deg) & (mag) & (SD) & (best) & & \\
 (1) & (2) & (3) & (4) & (5) & (6) & (7) & (8) & (9) & (10) & (11) & (12) & (13)\\
\hline
IC\,0676 & -1.3 & 93 & 170 & 58 $\pm$ 3 & - & 69 $\pm$ \hspace{4pt}6 & 69 $\pm$ \hspace{4pt}6 & -22.20 & - & - & C & 1 \\
IC\,0719 & -2.0 & 343 & 353 & 82 $\pm$ 1 & 79 $\pm$ 1 & 74 $\pm$ \hspace{4pt}5 & 74 $\pm$ \hspace{4pt}5 & -22.70 & x & x & C & 1 \\
IC\,1024 & -2.0 & 221 & 240 & 79 $\pm$ 1 & 72 $\pm$ 1 & - & 72 $\pm$ \hspace{4pt}1 & -21.70 & x & x & C & - \\
$^\star$IC\,2099 & -1.5 & 93 & - & 83 $\pm$ 1 & - & - & 83 $\pm$ \hspace{4pt}1 & -21.40 & ? & ? & - & - \\
NGC\,0524 & -1.2 & 312 & 320 & 28 $\pm$ 8 & 19 $\pm$ 7 & 44 $\pm$ 28 & 19 $\pm$ \hspace{4pt}7 & -24.70 & x & x & P & 2 \\
NGC\,1222 & -3.0 & 156 & 210 & 41 $\pm$ 5 & 41 $\pm$ 3 & - & 41 $\pm$ \hspace{4pt}3 & -22.70 & ? & - & C & 1 \\
NGC\,1266 & -2.1 & 187 & 180 & 47 $\pm$ 4 & 26 $\pm$ 5 & - & 26 $\pm$ \hspace{4pt}5 & -22.90 & - & - & C & 1 \\
NGC\,2685 & -1.0 & 156 & 220 & 70 $\pm$ 1 & 61 $\pm$ 1 & 70 $\pm$ 10 & 61 $\pm$ \hspace{4pt}1 & -22.80 & - & - & O & 4 \\
NGC\,2764 & -2.0 & 312 & 310 & 65 $\pm$ 2 & 65 $\pm$ 1 & 76 $\pm$ 15 & 65 $\pm$ \hspace{4pt}1 & -23.20 & x & x & C & 1 \\
NGC\,2768 & -4.4 & 322 & 360 & 36 $\pm$ 6 & - & - & 36 $\pm$ \hspace{4pt}6 & -24.70 & x & x & P & 2 \\
NGC\,2824 & -2.0 & 312 & 310 & 56 $\pm$ 3 & 61 $\pm$ 1 & - & 61 $\pm$ \hspace{4pt}1 & -22.90 & x & x & C & 1 \\
NGC\,3032 & -1.9 & 143 & 150 & 36 $\pm$ 6 & 35 $\pm$ 3 & 46 $\pm$ 11 & 35 $\pm$ \hspace{4pt}3 & -22.00 & x & x & B & 3 \\
NGC\,3182 & 0.4 & 218 & - & 39 $\pm$ 5 & 35 $\pm$ 3 & - & 35 $\pm$ \hspace{4pt}3 & -23.20 & x & x & - & - \\
NGC\,3489 & -1.2 & 292 & 240 & 62 $\pm$ 2 & - & 56 $\pm$ 15 & 62 $\pm$ \hspace{4pt}2 & -23.00 & ? & x & P & 2 \\
NGC\,3619 & -0.9 & 405 & - & 44 $\pm$ 4 & 48 $\pm$ 2 & - & 48 $\pm$ \hspace{4pt}2 & -23.50 & x & x & - & - \\
NGC\,3626 & -1.0 & 374 & 374 & 51 $\pm$ 3 & 61 $\pm$ 2 & 67 $\pm$ \hspace{4pt}5 & 67 $\pm$ \hspace{4pt}5 & -23.30 & x & x & C & 1 \\
NGC\,3665 & -2.1 & 624 & 630 & 90 $\pm$ 1 & 64 $\pm$ 1 & 74 $\pm$ 35 & 64 $\pm$ \hspace{4pt}1 & -24.90 & x & x & C & 1 \\
NGC\,4119 & -1.3 & 156 & 170 & 83 $\pm$ 1 & 67 $\pm$ 1 & 69 $\pm$ \hspace{4pt}3 & 69 $\pm$ \hspace{4pt}3 & -22.60 & ? & - & C & 1 \\
NGC\,4150 & -2.1 & 234 & 238 & 50 $\pm$ 3 & 54 $\pm$ 2 & 54 $\pm$ \hspace{4pt}3 & 54 $\pm$ \hspace{4pt}3 & -21.60 & x & x & B & 3 \\
$^\star$NGC\,4292 & -1.7 & 187 & 190 & 49 $\pm$ 3 & 50 $\pm$ 2 & 46 $\pm$ 14 & 50 $\pm$ \hspace{4pt}2 & -21.54 & ? & x & C & 1 \\
$^\star$NGC\,4309 & -1.5 & 124 & - & 69 $\pm$ 2 & 61 $\pm$ 1 & - & 61 $\pm$ \hspace{4pt}1 & -20.91 & - & - & - & - \\
NGC\,4324 & -0.9 & 218 & 360 & 73 $\pm$ 1 & 64 $\pm$ 1 & 62 $\pm$ \hspace{4pt}1 & 62 $\pm$ \hspace{4pt}1 & -22.60 & ? & x & C & 1 \\
NGC\,4429 & -1.1 & 499 & 532 & 68 $\pm$ 2 & 68 $\pm$ 1 & 60 $\pm$ 16 & 68 $\pm$ \hspace{4pt}1 & -24.30 & x & x & C & 1 \\
NGC\,4435 & -2.1 & 405 & 380 & 48 $\pm$ 4 & 52 $\pm$ 2 & - & 52 $\pm$ \hspace{4pt}2 & -23.80 & x & x & C & 1 \\
NGC\,4459 & -1.4 & 391 & 400 & 43 $\pm$ 4 & 46 $\pm$ 2 & 47 $\pm$ \hspace{4pt}2 & 47 $\pm$ \hspace{4pt}2 & -23.90 & x & x & B & 3 \\
NGC\,4477 & -1.9 & 204 & 260 & 36 $\pm$ 6 & 26 $\pm$ 5 & 38 $\pm$ \hspace{4pt}3 & 38 $\pm$ \hspace{4pt}3 & -23.70 & ? & x & P & 2 \\
NGC\,4526 & -1.9 & 655 & 663 & 82 $\pm$ 1 & 78 $\pm$ 1 & 64 $\pm$ \hspace{4pt}8 & 82 $\pm$ \hspace{4pt}1 & -24.60 & x & x & B & 3 \\
NGC\,4684 & -1.2 & 249 & - & 77 $\pm$ 1 & - & - & 77 $\pm$ \hspace{4pt}1 & -22.20 & ? & ? & - & - \\
NGC\,4694 & -2.0 & 93 & \hspace{4pt}65 & 69 $\pm$ 2 & - & - & 69 $\pm$ \hspace{4pt}2 & -22.10 & - & - & C & 1 \\
NGC\,4710 & -0.9 & 312 & 430 & 78 $\pm$ 1 & 88 $\pm$ 1 & 86 $\pm$ \hspace{4pt}6 & 86 $\pm$ \hspace{4pt}6 & -23.50 & x & x & C & 1 \\
NGC\,4753 & -1.4 & 530 & - & 61 $\pm$ 2 & 75 $\pm$ 1 & - & 75 $\pm$ \hspace{4pt}1 & -25.10 & x & x & - & 5 \\
NGC\,5173 & -4.9 & 187 & - & 24 $\pm$ 9 & - & - & 24 $\pm$ \hspace{4pt}9 & -22.90 & x & x & - & - \\
NGC\,5273 & -1.9 & 218 & - & 38 $\pm$ 5 & 38 $\pm$ 3 & - & 38 $\pm$ \hspace{4pt}3 & -22.40 & x & x & - & - \\
NGC\,5379 & -2.0 & 124 & 120 & 80 $\pm$ 1 & 64 $\pm$ 1 & - & 64 $\pm$ \hspace{4pt}1 & -22.10 & - & - & C & 1 \\
NGC\,6014 & -1.9 & 156 & 160 & 29 $\pm$ 7 & 22 $\pm$ 6 & 22 $\pm$ 10 & 22 $\pm$ \hspace{4pt}6 & -23.00 & x & x & C & 1 \\
NGC\,7465 & -1.9 & 158 & 180 & 56 $\pm$ 3 & 70 $\pm$ 1 & 58 $\pm$ \hspace{4pt}9 & 70 $\pm$ \hspace{4pt}1 & -22.80 & x & - & C & 1 \\
PGC\,29321 & 0.0 & 124 & - & 31 $\pm$ 7 & 38 $\pm$ 3 & - & 38 $\pm$ \hspace{4pt}3 & -21.60 & - & - & - & - \\
PGC\,56772 & -2.0 & 249 & - & 64 $\pm$ 2 & 57 $\pm$ 2 & - & 57 $\pm$ \hspace{4pt}2 & -22.00 & ? & ? & - & - \\
PGC\,58114 & -2.0 & 202 & 240 & 71 $\pm$ 1 & - & 76 $\pm$ 30 & 71 $\pm$ \hspace{4pt}1 & -21.60 & - & x & C & 1 \\
PGC\,61468 & 0.0 & 218 & - & 51 $\pm$ 3 & - & - & 51 $\pm$ \hspace{4pt}3 & -21.60 & ? & ? & - & - \\
UGC\,05408 & -3.3 & 156 & - & 31 $\pm$ 7 & - & - & 31 $\pm$ \hspace{4pt}7 & -21.90 & x & x & - & - \\
UGC\,06176 & -2.0 & 249 & 230 & 70 $\pm$ 2 & 68 $\pm$ 1 & - & 68 $\pm$ \hspace{4pt}1 & -22.60 & ? & - & C & 1 \\
UGC\,09519 & -1.9 & 187 & 210 & 47 $\pm$ 4 & 41 $\pm$ 3 & 63 $\pm$ \hspace{4pt}7 & 41 $\pm$ \hspace{4pt}3 & -22.10 & ? & x & C & 1 \\

\hline
\end{tabular*}
\label{table}
\\
\parbox[t]{\textwidth}{ \textit{Notes:} Column 1 lists the galaxy name. Galaxies with stars next to their name are early-types that were observed by Paper IV but are not included in the \atlas\ sample. Column 2 contains the morphological $T$-type from HyperLeda \citep{Paturel:2003p3431}. This morphology indicator was not used for the sample selection (see Paper I). Columns 3 and 4 list the velocity widths at 20\% of the peak flux, derived from single-dish and interferometric data, respectively. The error in these quantities was estimated as half the velocity width of an individual channel, 15 \kms. Column 5 contains the inclination derived from galaxy axial ratios. The quoted errors only take into account the error in the measurement of the optical axial ratio, as reported in NED. Column 6 contains the inclination derived from ellipse fitting to unsharp-masked dust images. The error is estimated to be 5\% on the measurement of both the minor and major axes.  Column 7 contains the inclinations estimated from tilted-ring fits to the interferometric datacubes, the errors on which are calculated as the maximum difference from the best value obtained when using only one half of the galaxy velocity field. Column 8 lists our adopted best inclinations, as described in Section \ref{bestinc}. Column 9 contains the $K_{\rm s}$-band magnitudes from 2MASS \citep{Skrutskie:2006p2829}, converted to absolute magnitudes using the distance to each galaxy adopted for the \atlas\ survey in Paper I. These distances are drawn preferentially from \cite{Mei:2007p3221} and \cite{Tonry:2001p3222}. We assign to the absolute magnitudes an error of $\pm$0.1 mag, taking into account distance uncertainties, which completely overwhelm errors caused by the lack of an internal extinction correction. Columns 10 and 11 list the galaxies with boxy CO profiles, as determined from visual inspection of the single-dish spectra (Column 9), and a combination of single-dish and interferometric data where available (Column 11). A question mark indicates that with the respective data we are unable to determine if the profile is boxy, and hence do not include it. The telescope used to obtain the interferometric data is indicated in Column 12, where C = CARMA, P = PdBI, B = BIMA and O = OVRO. Column 13 lists the relevant references for the interferometric observations, and in two cases (NGC\,2685 and NGC\,4753) references for the inclination measurements. (1) Alatalo et al., (in preparation) (2) \cite{Crocker2010}, (3) \cite{Young:2008p788}, (4) \cite{Jozsa:2009p3232} and (5) \cite{SteimanCameron:1992p2507}.}
\end{table*}

\section{Comparison with other velocity measures}
\subsection{Model circular velocity curves}
\label{modelcurves}

As alluded to in Section \ref{velwidths}, it has yet to be established for a large sample of early-type galaxies that single-dish CO is a good tracer of the circular velocity. 
One would like to test this
, and attempt to quantify the biases introduced by using the profile shape as a proxy for reaching beyond the peak of the circular velocity curve. As \cite{Williams2010} discuss, one must be careful when comparing velocity measures derived from different methods, as they are likely to have significant systematic differences. However, it is important to ascertain that the CO line-widths vary systematically, in a similar way to other rotation measures. 

The gravitational potential of ETGs can also be significantly different from that of spirals. ETGs often have circular velocity profiles with high peaks at small radii that decline before flattening out. The W$_{20}$ line-width is, in general, more sensitive to the peak velocity of the molecular gas, rather than the flat part of the rotation curve, unless the gas disk is sufficiently extended. This would lead to an overestimation of the velocity width. This problem has been discussed in detail by \cite{Noordermeer:2007p3078}, who find that using the \hi\ peak velocity in massive early-type spirals results in an offset TFR. They report that the asymptotic velocity is a better measure of the total potential, and results in spirals of all masses lying on a single TFR. 

An estimate of the amount by which we may be overestimating the velocity widths can be obtained using the circular velocity curves produced by \cite{Williams:2009p2681} from stellar dynamical modeling of their S0 galaxies. The median value of the circular velocity curve after the peak out to the maximum extent of the data, where the model is well constrained, is adopted as a measure of the velocity of its flat part (in the case of \cite{Williams:2009p2681} their data extends to $\approx$3 Re). When tested on the S0 circular velocity curves from \cite{Williams:2009p2681}, this median performs well in picking out the velocity of the flat section.
The mean ratio of the peak velocity to the flat velocity is 1.15, or 0.06 dex.  When transformed to an offset in luminosity using the parameters of the best-fit S0 TFR of \cite{Williams2010}, this effect corresponds to an offset of upto 0.5 mag at $K_{\rm{s}}$-band. This is approximately half of the $\approx$1 mag offset observed by various authors between the spiral and ETGs, and is thus significant if not properly accounted for.

In order to test the assumption that our ETGs with boxy profiles have gas distributions that reach beyond the peak of the circular velocity curve, and to see if our line-widths are overly sensitive to the peak velocity, like \cite{Williams:2009p2681,Williams2010} we utilize the axisymmetric Jeans anisotropic dynamical modeling (JAM) method described in \cite{Cappellari:2008p2773}. Some examples of the approach, using SAURON integral-field kinematics \citep{Emsellem:2004p1497} as done here, are presented in \cite{Scott:2009p3218}. For the \atlas\ survey a Multi-Gaussian expansion \citep[MGE;][]{Emsellem:1994p723} was fitted to the SDSS \citep{Abazajian:2009p3430} or INT photometry (Scott et al. in preparation). The MGEs were then used to construct JAM models for all the 260 \atlas\ galaxies \citep[see][a]{Cappellari:2010p3429} which were fitted to the SAURON stellar kinematics (Paper I).
The models have three free parameters, the inclination ($i_{\rm JAM}$), the mass-to-light ratio (M/L) assumed to be spatially constant, and the anisotropy $\beta_z=1-\sigma_z^2/\sigma_R^2$, which is also assumed to be spatially constant. From each mass model we have calculated the predicted circular velocity curve in the plane of the galaxy. It is worth bearing in mind that, where the CO is misaligned from the plane of the galaxy, we are thus making implicit assumptions about the symmetry of the matter distribution. As these models include no dark matter, the circular velocity often declines at large radii, and one must be careful to measure the circular velocity at a suitable radius (i.e. where we have constraining data) when comparing to the observations. 

Analysis of the \atlas\ interferometric observations, presented in Davis et al., (in preparation), suggests that the average radial extent of CO discs in early-type galaxies is around one optical effective radius (the radius encompassing half of the light), hence the circular velocity measured at one effective radius (denoted V$_{\rm JAM,Re}$) is a sensible quantity to use for comparison. The effective radius used here is a combination of the values from 2MASS \citep{Skrutskie:2006p2829} and RC3 \citep{deVaucouleurs:1991p2406}, as described in Paper I. For galaxies where interferometric observations are available, we have also calculated the model circular velocity at the maximum extent of the CO, and denote this V$_{\rm JAM,R_{CO}}$. Both values can be found in Table \ref{tableJAM}.

\begin{figure}
\begin{center}
\includegraphics[scale=0.7,angle=0,clip,trim=0.5cm 0cm 0.0cm 1cm]{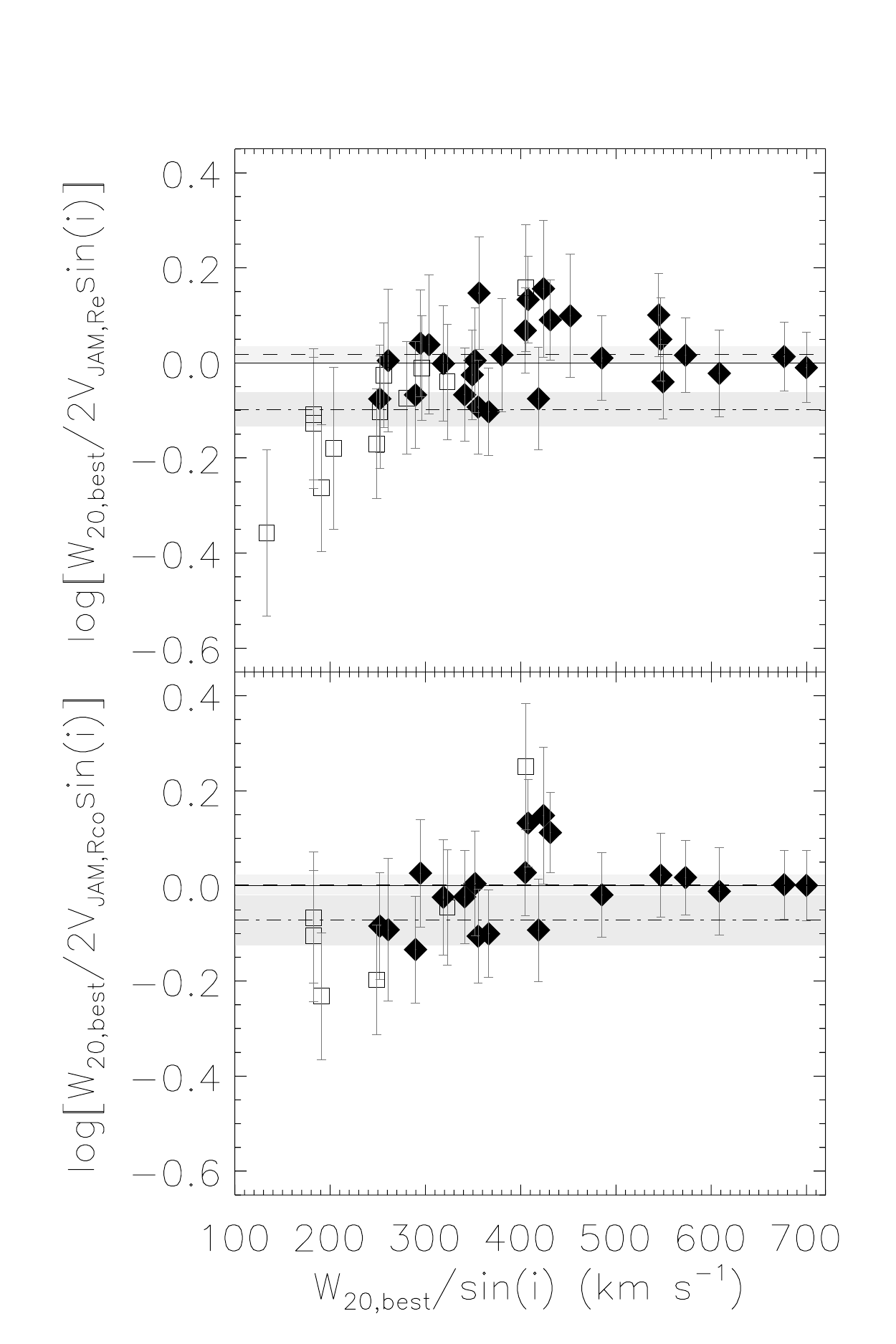}
\end{center}
\caption{\small Difference between our best CO line-widths de-projected using the best inclinations (as described in Section \ref{bestinc}) and twice the JAM model circular velocities, measured at one effective radius (top) and the radius of maximum CO extent, or 12\arcsec\, whichever is smaller (bottom), plotted against the best inclination de-projected CO velocity width. Galaxies with boxy profiles are plotted as filled diamonds, while the others are shown as open squares. The solid line indicates where the JAM and CO velocities are identical. The dashed line and dot-dashed line are fitted offsets from zero, for the boxy galaxies and the other galaxies, respectively. The errors on the fits are indicated as shaded regions around the best fit lines.}
\label{coworks}
\end{figure}

We compare the two model circular velocity measures described above with the measured CO line-widths in Figure \ref{coworks}. The line-widths of the galaxies with boxy CO profiles 
correlate well with the circular velocities measured at both one effective radius (R$_{\rm e}$) and the radius of maximum CO extent (R$_{\rm CO}$). The offset from V$_{\rm JAM,Re}$ is 0.01 $\pm$ 0.02 dex, the data points displaying an RMS scatter of 0.1 dex. The galaxies that do not have boxy CO profiles have line-widths systematically smaller than those predicted from the models in most cases, suggesting that their CO has been correctly identified as not reaching beyond the peak of the rotation curve.
The comparison with V$_{\rm JAM,Rco}$, albeit with smaller number statistics, shows that the difference between the model and the CO line-widths is again consistent with the models, with a mean offset of 0.002 $\pm$ 0.02 dex and a RMS scatter of 0.1 dex. There is no systematic behaviour in the residuals for the boxy galaxies in either plot. The galaxies that do not have boxy CO profiles are closer to the JAM model predictions, but do not lie on the one to one relation as expected, with a mean offset of -0.07$\pm$0.05 dex. This offset is likely because the CO is these objects is very compact, and beam effects mean the CO size is overestimated. The JAM model velocities have thus been estimated at the wrong radius.
The one galaxy which does not have a boxy profile, but has a line-width larger than the predicted circular velocity, is NGC\,1266, recently discovered to have a kiloparsec-scale molecular outflow \citep{ngc1266}. 

To check if our line-widths are overly-sensitive to the peak of the rotation curve, we also extract from the JAM models the maximum of the circular velocity curve, denoted V$_{\rm JAM,max}$, and the median value of the circular velocity curve, measured between the peak and the maximum extent of the \sauron\ data ($\approx$1R$_{\rm e}$), denoted V$_{\rm JAM,>peak}$. As discussed above, defining V$_{\rm JAM,>peak}$ in this way picks out the value of the flat part of the rotation curve well in tests based on the data of \cite{Williams:2009p2681}, which extend to $\approx$3 Re. Our data is only constrained out to around $\approx$1R$_{\rm e}$, but the rotation curves do not drop wildly in this region. It should however be noted that this measure may not be the same as the velocity measured at very large radii. The values of V$_{\rm JAM,max}$ and V$_{\rm JAM,>peak}$ can be found in Table \ref{tableJAM}.

\begin{figure}
\begin{center}
\includegraphics[scale=0.7,angle=0,clip,trim=0.5cm 0cm 0.0cm 1cm]{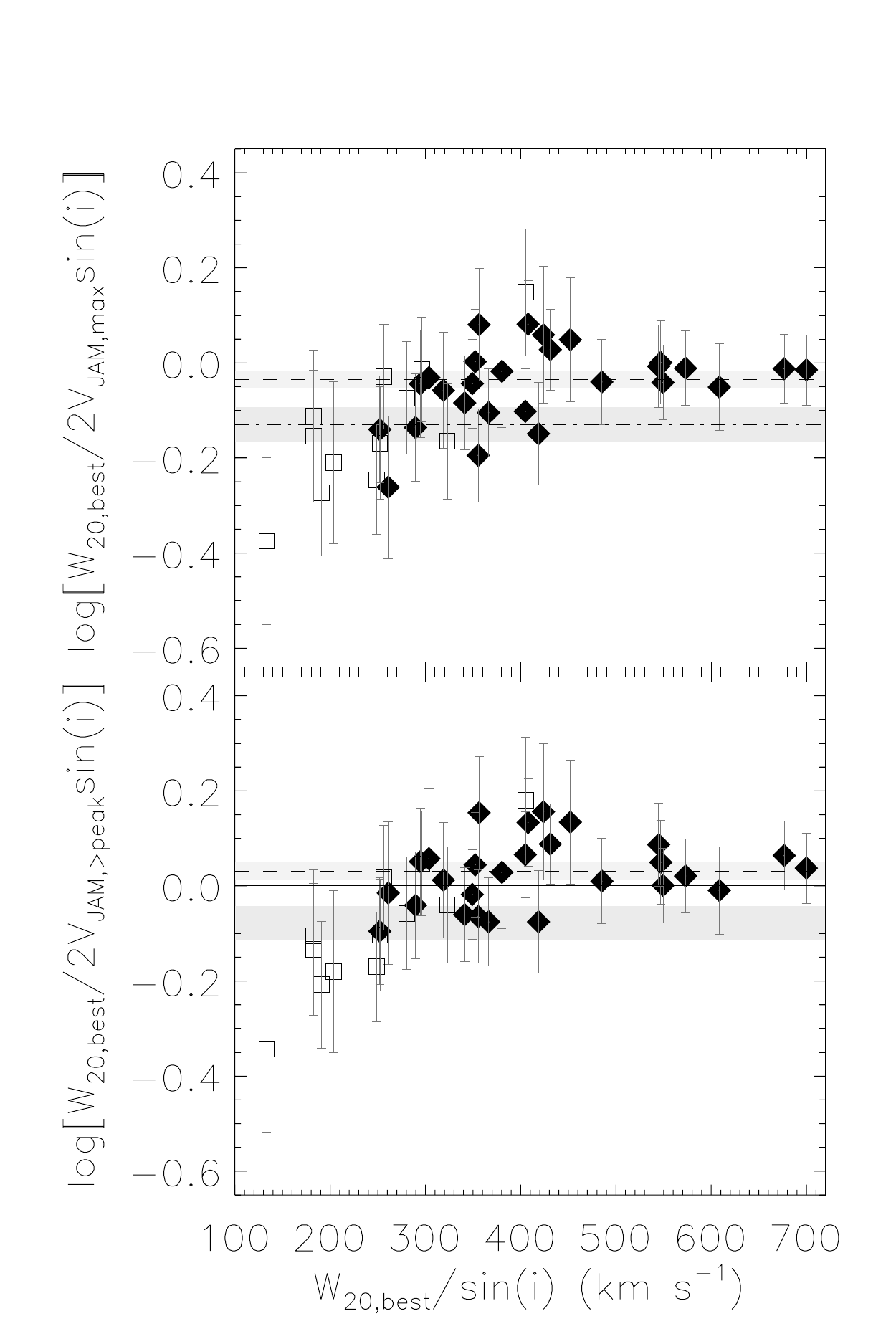}
\end{center}
\caption{\small Same as Figure \ref{coworks}, but for JAM model circular velocities measured at the maximum (top) and using the median velocity after the maximum (bottom).}
\label{flatnotmax}
\end{figure}

The top panel of Figure \ref{flatnotmax} shows a comparison between the observed CO line-widths, de-projected using dust inclinations (as described in Section \ref{galdust}), and V$_{\rm JAM,max}$. The CO line-widths are systematically smaller than the maximum of the circular velocity curve. The median peak of the JAM circular velocity curves is at 8.5\arcsec, well within the single-dish beam, and hence we do not expect systematically missed emission to cause this trend. The mean offset  from the maximum value is -0.03 $\pm$ 0.02 dex, broadly similar to the offset between the maximum and the flat part of the circular velocity curve expected from the results of \cite{Williams:2009p2681}. The RMS scatter around this value is 0.09 dex. This suggests that we are not overly sensitive to the peak of the circular velocity curve. The galaxies which do not have boxy profiles are again offset to much lower velocities. The only galaxy whose line-width is inconsistent with the maximum value of its predicted circular velocity curve is again NGC\,1266.

The bottom panel of Figure \ref{flatnotmax} compares the observed CO line-widths, de-projected using dust inclinations, with V$_{\rm JAM,>peak}$. The CO measurements for galaxies with double horned profiles are, within the errors, consistent with the V$_{\rm JAM,>peak}$ measure. The mean offset from this measure is 0.02 $\pm$ 0.02 dex, with an RMS deviation of 0.1 dex. 
The galaxies that were not identified as boxy are on average offset from V$_{\rm JAM,>peak}$ by -0.1$\pm$0.04 dex. There are several galaxies that do not have a boxy profile, but are consistent with rotating at V$_{\rm JAM,>peak}$. These galaxies are NGC\,4684, PGC\,56772 and PGC\,61468, all of which have profiles which were hard to classify. If interferometric observations were available we would hence expect to find that the gas does indeed reach beyond the peak of the galaxy rotation curve in these systems.

Figures \ref{coworks} and \ref{flatnotmax} demonstrate that, if one selects galaxies with boxy profiles, W$_{20}$ velocity widths provide an good estimate of the circular velocity after the inner peak of the rotation curve, whether estimated as V$_{\rm JAM,Re}$ or V$_{\rm JAM,>peak}$.  It is worth mentioning that, for our sample, a simple cut excluding galaxies with CO linewidths $\ltsimeq$250\kms\ would have a similar effect as selecting by profile shape, removing most of the outliers. Applying an arbitrary cut in velocity, the most important TF parameter, could bias the derived TFR, especially for a sample selected in a different way, and hence we prefer selection by profile shape in this work.

\begin{table}
\caption{Parameters extracted from the JAM model circular velocity curves for the \atlas\ early-type CO-rich galaxies used in this paper. }
\begin{tabular*}{0.5\textwidth}{@{\extracolsep{\fill}}r r r r r} 
\hline

Galaxy & V$_{\rm JAM,Re}$ & V$_{\rm JAM,Rco}$ & V$_{\rm JAM,max}$ & V$_{\rm JAM,>peak}$\\
 & (\kms) & (\kms) & (\kms) & (\kms)\\
 (1) & (2) & (3) & (4) & (5)\\
\hline
IC\,0676 & 117 & 116 & 118 & 116\\
IC\,0719 & 232 & 232 & 233 & 218\\
IC\,1024 & 150 & 163 & 174 & 157\\
IC\,2099 & - & - & - & -\\
NGC\,0524 & 337 & 362 & 368 & 337\\
NGC\,1222 & 177 & 177 & 236 & 177\\
NGC\,1266 & 141 & 116 & 144 & 134\\
NGC\,2685 & 160 & - & 186 & 160\\
NGC\,2764 & 199 & 190 & 207 & 196\\
NGC\,2768 & 320 & 325 & 342 & 311\\
NGC\,2824 & 220 & 220 & 278 & 206\\
NGC\,3032 & 129 & 158 & 238 & 135\\
NGC\,3182 & 183 & - & 198 & 178\\
NGC\,3489 & 169 & 197 & 198 & 159\\
NGC\,3619 & 216 & - & 277 & 223\\
NGC\,3626 & 173 & 187 & 256 & 174\\
NGC\,3665 & 358 & 354 & 362 & 321\\
NGC\,4150 & 134 & 141 & 163 & 131\\
NGC\,4270 & 185 & - & 193 & 182\\
NGC\,4292 & - & - & - & -\\
NGC\,4309 & - & - & - & -\\
NGC\,4324 & 150 & 152 & 169 & 150\\
NGC\,4429 & 276 & 276 & 294 & 273\\
NGC\,4435 & 237 & 263 & 266 & 237\\
NGC\,4459 & 244 & 261 & 273 & 244\\
NGC\,4477 & 249 & 278 & 295 & 249\\
NGC\,4526 & 328 & 338 & 348 & 292\\
NGC\,4684 & 136 & - & 137 & 123\\
NGC\,4694 & 96 & - & 99 & 93\\
NGC\,4710 & 175 & 161 & 202 & 176\\
NGC\,4753 & 301 & - & 302 & 274\\
NGC\,5173 & 180 & - & 202 & 166\\
NGC\,5273 & 127 & - & 148 & 125\\
NGC\,5379 & 152 & - & 158 & 147\\
NGC\,6014 & 148 & 149 & 185 & 148\\
PGC\,29321 & 154 & - & 165 & 154\\
PGC\,56772 & 152 & - & 153 & 133\\
PGC\,58114 & - & - & - & -\\
PGC\,61468 & 166 & - & 166 & 160\\
UGC\,05408 & 139 & - & 163 & 133\\
UGC\,06176 & 184 & 196 & 219 & 184\\
UGC\,09519 & 160 & 165 & 182 & 155\\
\hline
\end{tabular*}
\label{tableJAM}
\textit{Notes:} This table contains the circular velocities derived from the JAM models, extracted at different radii. Galaxies that were removed from the \atlas\ sample have no JAM model available. Column 2 contains the circular velocities extracted at one effective radius. Column 3 contains the circular velocities extracted at the maximum radial extent of the CO, or 12\arcsec, whichever is smaller. Galaxies without interferometric observations are not included. Column 4 contains the maximum circular velocity. Column 5 contains the median of velocities after the peak in the circular velocity curve out to the maximum radius at which the model is constrained by the observational data. Some galaxies had insufficient data to constrain such a median between these two points, and are excluded. The error on these JAM velocities is estimated to be $\approx$8\%. This is a combination of an average 5\% error in the observed stellar velocity dispersion constraining the models, and model errors of $\approx$6\% \citep{Cappellari:2006p1498}. This does not include errors in inclination, which are harder to quantify.
\end{table}

\subsection{\hi\ line-widths}
\cite{Morganti:2006p1934} and \cite{Williams2010} have shown that in some cases \hi\ line-widths are 
not a good measure of the circular velocity for early-type galaxies. We examine this here by retrieving archival \hi\ single-dish line-widths measured at 20\% of the peak flux from HyperLEDA \citep[][denoted W$_{20,\rm HI}$]{Paturel:2003p3431}. 

\begin{figure}
\begin{center}

\includegraphics[scale=0.55,angle=0,clip,trim=0.5cm 0cm 0.5cm 1cm]{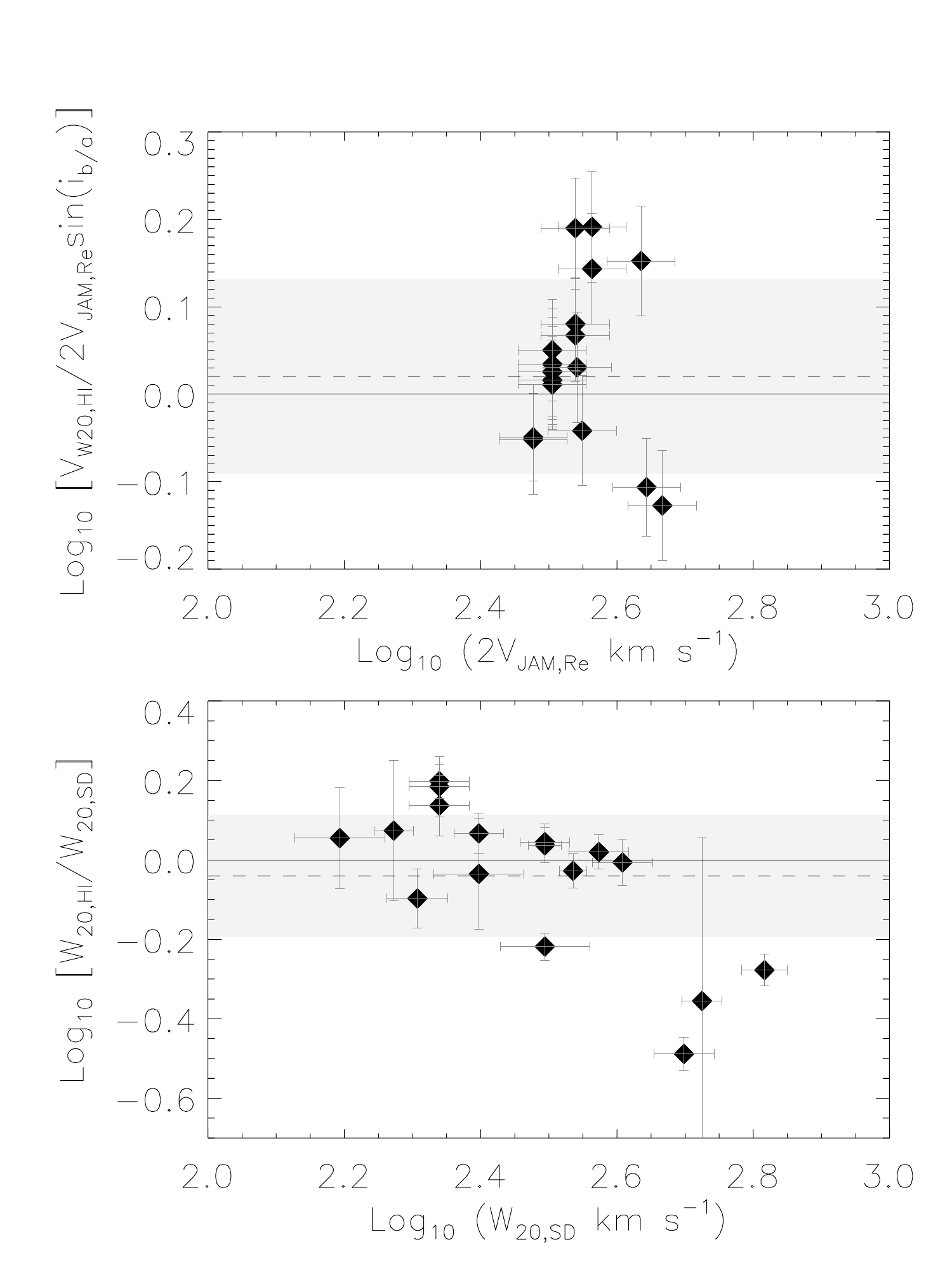} 
\end{center}
\caption{\small \textit{Top:} Difference between the deprojected \hi\ line-width and twice the JAM model circular velocity measured at one effective radius, for galaxies with archival \hi\ data in LEDA, as a function of the model velocities. The solid line indicates where the JAM and \hi\ velocities are identical. The dashed line is the mean difference, with the RMS scatter indicated in grey. \textit{Bottom:} As for the top panel, but comparing observed CO single-dish and \hi\ line-widths, for galaxies with boxy CO profiles only. }
\label{hidoesntwork}
\end{figure}

In Figure \ref{hidoesntwork}, we compare the single-dish \hi\ line-widths 
to twice the projected JAM circular velocities measured at R$_{\rm e}$ (top panel) and the observed CO line-widths (bottom panel). All galaxies from our sample with archival \hi\ data at the time of retrieval are included.
We find that although the average difference between the \hi\ line-width and the JAM circular velocities is small (0.02 dex), the scatter is large (RMS scatter of $\approx$0.11 dex), and the single-dish \hi\ line-widths do not correlate statistically with the galaxy luminosity. 
The same is true when one compares with observed CO line-widths, with a larger scatter ($\approx$0.15 dex). There is also a systematic trend in the residuals in the bottom panel of Fig. \ref{hidoesntwork}, where the galaxies with high CO rotation velocities, which are likely of high masses, have systematically smaller \hi\ line-widths.  

Figure \ref{hidoesntwork} suggests that some of the detected \hi\ sources, at high CO velocities (and hence likely at high mass), are either not related to the galaxy under study or that the \hi\ has been kinematically disturbed. This is consistent with the results of \cite{Morganti:2006p1934}, \cite{Williams2010} and Serra et al., (in preparation). 

Single-dish \hi\ Line-widths in the LEDA database are however known to suffer from many issues, including major problems with source confusion. Outer HI disks are also usually warped with respect to the inner regions, and hence a simple inclination estimates from optical images will be insufficient. Interferometric \hi\ data allows one to identify relaxed disks, and in these rotation and the TFR can be studied \citep[e.g.][]{Morganti:2006p1934,Oosterloo:2007p3288,Weijmans:2008p3286,Oosterloo:2010p3287}. To further investigate the TFR, and the disk-halo consipracy in these galaxies, a future paper in this series will study the \hi\ TFR of ETGs at large radii using superior interferometric \hi\ data.

\section{Results}

Using the data presented in Table \ref{table}, we construct a series of Tully-Fisher relations to explore the effect of the different inclination and velocity measurements, defined in order of increasing complexity. The results are shown in Figure \ref{tf_rels}. The general form of the $K_{\rm s}$ band Tully-Fisher relation we have adopted is

\begin{equation}
\label{fiteq}
{\rm M}_{K} =  \rm a\,[{\rm log}_{10}\left(\frac{W_{20}}{km\ s^{-1}}\right) - 2.6]+b,
\end{equation}
where a is the slope and b is the zero-point of the relation. 

The package {\tt MPFIT} \citep{Markwardt:2009p2588} was used to fit the inverse of Equation \ref{fiteq} (regressing the observed rotation velocities) with the addition of an intrinsic scatter, which we iteratively adjust to ensure a reduced $\chi^2\approx1$. Full details of the fitting procedure can be found in \cite{Williams2010}.

Each plot in Figure \ref{tf_rels} shows two best-fit lines. For the first we allow both the intercept and slope to vary, while for the second we fix the slope to that found by  \cite{Tully:2000p2442}, a = -8.78. For reference we also plot the $K$-band spiral galaxy TFR of \cite{Tully:2000p2442} (see Equation \ref{tully2000eq}), which was constructed using \hi\ line-widths of spiral galaxies in 12 well separated clusters, and the $K_{\rm s}$-band S0 TFR from \cite{Williams2010} (see Equation \ref{mikeeq}), which was derived from Jeans modeling of major-axis stellar kinematics from 14 edge-on S0 galaxies:
\begin{equation}
\label{tully2000eq}
{\rm M}_{K} =  \rm -8.78\,[{\rm \log}_{10}\left(\frac{W_{20}}{km\ s^{-1}}\right) - 2.5]-23.17 ,
\end{equation}

\begin{equation}
\label{mikeeq}
{\rm M}_{K\rm{s}} =  \rm -8.56\,[{\rm \log}_{10}\left(\frac{W_{20}}{km\ s^{-1}}\right) - 2.4]-24.02\,.
\end{equation}
\subsection{CO Tully-Fisher relations}
\subsubsection{Inclinations from galaxy axial ratios}

The CO TFR de-projected using $i_{\rm b/a}$ is presented in the top two panels of Figure \ref{tf_rels}. The parameters of the best-fit relations are listed in Table \ref{fittable}. The top-left panel shows all galaxies, with open squares for those which do not show a boxy CO profile or where the profile shape is unclear, and solid triangles for those with clear boxy profiles. Only a loose correlation is observed, with clear outliers, almost all of which are galaxies without double horned profiles. A much tighter correlation is observed once one removes galaxies with non-boxy profiles, as shown in the top-right panel. This is physically motivated, as we believe the gas in these galaxies does not reach beyond the peak of the rotation curve. Once the non-doubled-horned galaxies are removed both best-fit lines agree within the errors with the TFR of \cite{Williams2010}, providing strong \textit{a posteriori} evidence that the CO in these galaxies does trace the circular velocity in a way that is consistent with dynamical models. 
The remaining outlier is NGC\,7465 which we will discuss in more detail below. If this outlier is removed the best fit relation follows that of \cite{Williams2010} even more closely. The intrinsic scatter of the best fit relation with a free, and a constrained, slope is respectively 0.57 and 0.63 mag, with total RMS scatters of 0.69 and 0.78 mag. 

\subsubsection{Inclinations from dust axial ratios}
In a similar way we can construct a TFR using the inclinations derived from fitting ellipses to the dust distributions highlighted in unsharp-masked images.
The CO TFR deprojected using $i_{\rm dust}$ is presented in the middle-left panel of Figure \ref{tf_rels}. Once again, within the quoted errors, the best-fit relations are consistent with the result of \cite{Williams2010}. The parameters of the best-fit relations are listed in Table \ref{fittable}. The intrinsic scatter of the best-fit relation with a free, and a constrained, slope is 0.65 and 0.78 mag respectively, but with a total scatter of 0.75 and 0.94 mag, respectively. The outlier is NGC\,7465 - without this galaxies the observed intrinsic scatter in the unconstrained fit decreases to 0.46 mag, with a total scatter of 0.6 mag, confirming the visual impression that the correlation using dust inclinations is tighter. Removing this galaxy also improves the agreement between our best-fit slope with that found by \cite{Williams2010}.

\subsubsection{Inclination from molecular gas modeling}

\label{tiltringsec}
Using the inclinations from the tilted-ring models and the total velocity widths calculated from the interferometric data, we can construct yet another 
 TFR. First we do this naively, using only the single-dish boxy classifications. This results in the middle-right panel of  Figure \ref{tf_rels}. Some of the interferometric observations have insufficient spatial resolution to strongly constrain the inclination of the molecular gas, and this is reflected in the error bars. As can clearly be seen in this panel, the best-fit relation is only marginally consistent with that of \cite{Williams2010}. The intrinsic scatter of the best-fit relation with a free, and a constrained, slope is 0.43 and 0.62 mag, with a total scatter of 0.60 and 0.82 mag, respectively. 

The interferometric data, at least in some cases, allow us to further refine our technique for identifying galaxies that have reached beyond the peak of the circular velocity curve, by identifying nuclear rings and finding galaxies where a turnover can be seen in the position-velocity diagram. Interferometric observations are especially powerful in this regard, as the extent of the CO can be compared to the expected circular velocity curve from dynamical models. For example, NGC\,7465, the obvious outlier in the previous plots, features a misaligned circum-nuclear CO distribution, which has a double horned profile, but has no turnover in its velocity field, and is extremely unlikely to reach beyond the peak of the galaxy rotation curve.

Aperture synthesis observations also allow us to identify cases where a galaxy is misclassified using the single-dish data because a pointing error and/or extended molecular gas results in only some smaller portion of the total velocity width being detected. Such cases include NGC\,4324, NGC\,4477 and PGC\,58114. In a few cases where the profile shape was uncertain (due to low signal to noise single dish spectra) the interferometric data reveals that the gas does extend beyond the peak of the rotation curve, allowing us to include them. This is the case for NGC\,3489, NGC\,4292 and UGC\,09519.

The TFR in the middle-right panel of Figure \ref{tf_rels} can thus be improved using the full knowledge gained from the interferometric data (discussed above), and this results in the TFR shown in the bottom-left panel of Figure \ref{tf_rels}. The best-fit TFR using this additional information agrees with that of \cite{Williams2010}. The parameters of the best-fit relations are again listed in Table \ref{fittable}. The intrinsic scatter of the best-fit relation with a free, and a constrained, slope is 0.32 and 0.33 mag, with total a scatter of 0.54 and 0.58 mag, respectively.

\subsubsection{Combining interferometric and single-dish data}
\label{best}
\begin{figure*}
\begin{center}
\includegraphics[scale=1.13,angle=0,clip,trim=0.5cm 0cm 1cm 1cm]{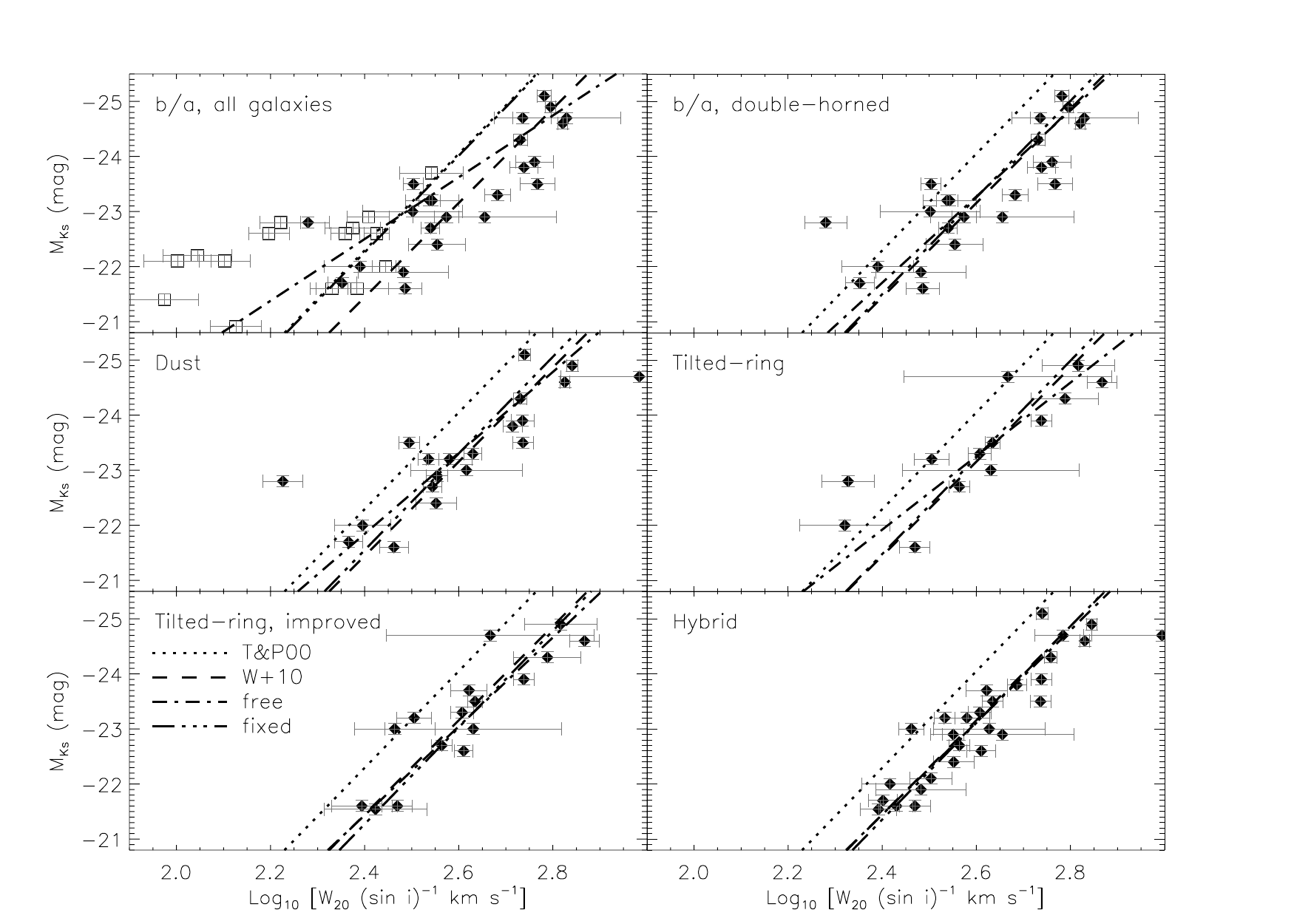}
\end{center}
\caption{\small CO TFRs constructed using various inclination and line-width measurements. The correlations shown in each plot and indicated in the legend are the spiral galaxy TFR of \protect \cite{Tully:2000p2442}, derived from \hi\ line-widths (short dashes), and the S0 TFR of \protect \cite{Williams2010}, derived from Jeans modeling (long dashes). Also plotted are the best-fit relations for our data, of the form given in Equation \protect \ref{fiteq}, for both an unconstrained fit (dot-dashed) and a fit where the slope is fixed to that found by \protect \cite{Tully:2000p2442} (triple dot-dashed). The best-fit parameters are listed in Table \ref{fittable}.
\textit{Top-left:} Inclinations derived from galaxy optical axial ratios (see Section \ref{galaxial}). CO detections with boxy profiles are shown as filled diamonds, galaxies with non-boxy profiles and uncertain classifications are shown as open squares. \textit{Top-right:} As top-left, but galaxies with non-boxy profiles are removed. The outlier is NGC\,7465. \textit{Middle-left:} As top-right, but using inclinations from unsharp-masked dust images (see Section \ref{galdust}). The outlier is again NGC\,7465. \textit{Middle-right:} As top-right, but using interferometric W$_{20}$ values and inclinations derived from tilted-ring fits to the interferometric data (see Section \ref{inter}). \textit{Bottom-left:} As middle-right, but using the interferometric data to identify galaxies with CO which reaches beyond the peak of the galaxy rotation curve. \textit{Bottom-right:} A hybrid plot using the best inclinations and velocity widths as defined in Section \protect \ref{bestinc}. Both best-fit lines overlay the relation of Williams et al.}
\label{tf_rels}
\end{figure*}

Finally, we can combine our single-dish and interferometric observations, using the hybrid inclinations and velocities described in Section \ref{bestinc}.
These choices lead to the TFR shown in the bottom-right panel of Figure \ref{tf_rels}. The parameters of the best-fit TF relations are listed in Table \ref{fittable}. The best-fit relations closely agree with the result of \cite{Williams2010}, and the fits with a free and a constrained slope have an intrinsic scatter of 0.36 and 0.37 mag, with a total scatter of 0.54 and 0.57 mag, respectively.  The intrinsic scatter found in this relation is similar to that found by a large study of the intrinsic scatter of the spiral galaxy TFR \citep[$\sigma_{\rm int} \approx 0.4$ mag at bands between $g$ and $z$;][]{Pizagno:2007p2836}, and only very slightly larger than that found by \cite{Williams2010} ($\sigma_{\rm int} \approx 0.3$ mag at $K_{\rm{s}}$-band).
The small total scatter reflects both the better number statistics and the use of the best possible measures of rotation and inclination. 

Using this hybrid relation we can investigate if the residuals have any systematic behavior with respect to the properties of the hosts. We find that the residuals are statistically uncorrelated with the galaxy $K_{\rm{s}}$-band luminosity \citep{Skrutskie:2006p2829}, optical colour \citep[$g$-$r$;][Scott et al., in preparation]{AdelmanMcCarthy:2008p2834}, velocity dispersion, galaxy total dynamical mass (Cappellari et al. in preparation) and spin parameter $\lambda_{R}$ \citep[][Paper III]{Emsellem2010}

\begin{table*}
\parbox[t]{0.9 \textwidth}{\caption{Best-fit parameters of the TFRs shown in Figure {\protect \ref{tf_rels}}\label{fittable}.}}

\begin{tabular*}{0.9\textwidth}{@{\extracolsep{\fill}}c c c c c c c c} 
\hline
Inc. method & a (mag)& b (mag) & $\sigma_{\rm int}$ (mag) & $\sigma_{\rm total}$ (mag) & W$_{\rm 20,sd}$ &  W$_{\rm 20,inter}$ & Selection\\
 (1) &(2) &(3) & (4) & (5) & (6) & (7) & (8)\\
\hline
Stellar $b$/$a$ & -5.61 $\pm$ 0.71 & -23.62 $\pm$ 0.16 & 0.75 & 0.82 & x & -& none\\
Stellar $b$/$a$ & - & -24.01 $\pm$ 0.23 & 1.30 & 1.42 & x & -& none\\
Stellar $b$/$a$ & -7.82 $\pm$ 1.12 & -23.27 $\pm$ 0.15 & 0.60 & 0.69 & x & -& SD\\
Stellar $b$/$a$ & - & -23.25 $\pm$ 0.16 & 0.66 & 0.78 & x & -& SD\\
Dust & -7.36 $\pm$ 1.19 & -23.31 $\pm$ 0.16 & 0.66 & 0.76 & x & -& SD\\
Dust & - & -23.30 $\pm$ 0.19 & 0.79 & 0.94 & x & -& SD\\
Tilted-ring & -6.68 $\pm$ 1.21 & -23.26 $\pm$ 0.17 & 0.48 & 0.63 & - & x& SD\\
Tilted-ring & - & -23.23 $\pm$ 0.23 & 0.65 & 0.86 & - & x& SD\\
Tilted-ring & -8.07 $\pm$ 1.10 & -23.05 $\pm$ 0.12 & 0.32 & 0.54 & - & x& Inter\\
Tilted-ring & - & -23.03 $\pm$ 0.13 & 0.33 & 0.58 & - & x& Inter\\
Hybrid & -8.38 $\pm$ 0.70 & -23.12 $\pm$ 0.09 & 0.36 & 0.54 & x & x& Inter+SD\\
Hybrid & - & -23.11 $\pm$ 0.09 & 0.37 & 0.57 & x & x& Inter+SD\\
\hline

\end{tabular*}
\\
\parbox[t]{0.9 \textwidth}{ \textit{Notes:} Column 1 describes the inclination method used to deproject the measured line-width. Columns 2 and 3 show the best-fit slope and zero-point, respectively, of the Tully-Fisher relation of the form ${M}_{K\rm{s}} =  \rm a[{\rm log}_{10}(W_{20}/km\,s^{-1}) - 2.6]+b$. Where Column 2 is blank the fit was constrained to have the same slope as that found by \protect \cite{Tully:2000p2442}, a=-8.78. Column 4 shows the intrinsic scatter ($\sigma_{\rm int}$) required by the fit to produce a reduced $\chi^2\approx1$. Column 5 shows the total RMS scatter around the best-fit relation. A cross in Column 6 denotes that the velocity widths used are from single-dish data, while a cross in Column 7 denotes that interferometric line-widths were used. Column 8 lists the selection method used to find galaxies that have gas beyond the peak of the circular velocity curve. `SD' means the galaxies single-dish profiles are boxy, while `Inter' means that interferometric data were used to determine if the gas reaches beyond the peak of the circular velocity curve. The hybrid relations use both single-dish and interferometric data, as described in Section \protect \ref{bestinc}.}
\end{table*}

\section{Discussion}

\subsection{\co\ as a tracer of galactic potentials}

The results presented in Section \ref{modelcurves} show that \co\ emission accurately traces the potential of early-type galaxies, if one has a suitable method of identifying which galaxies have extended molecular discs. 
Our work demonstrates that selecting galaxies via profile shape is robust and does not systematically affect the measured rotation velocity. The main possible contaminant introduced by this selection is galaxies that have molecular rings, which would display a boxy profile without necessarily rotating at \vmax. Circumnuclear rings pose the greatest problem, as they are very unlikely to rotate at \vmax, unlike inner and outer rings which are located at resonance points within the flat part of the galaxy circular velocity curve \citep[e.g.][]{Buta:1986p3227}. The number of galaxies with circumnuclear rings in our sample is likely to be small, as their presence in significant numbers would bias the best-fit line in Figure \ref{coworks} to small values. This conclusion is also supported by the interferometric survey of Alatalo et al., (in preparation), which suggests that only a small percentage of local early-type galaxies have their molecular gas restricted to the circumnuclear region. 

One should also remember that the circular velocity curves of ETGs rise very quickly, normally peaking within one effective radius \citep[e.g.][]{Williams:2009p2681}.  
This allows the molecular distribution to be compact, but still have enough gas beyond the peak of the galaxy circular velocity curve so that the linewidth is a good proxy of the circular velocity in that region. The good agreement between our derived TFR and that of \cite{Williams2010} suggests that the disk-halo conspiracy holds for our early-type galaxies, at least out to 3R$_{\rm e}$. If this conspiracy holds even further out into the dark matter dominated parts of these galaxies \citep[which some authors suggest it does not, e.g.][]{Dutton2010} will be considered in a future paper in this series.

\subsection{Inclination measures}
\label{discussinc}
The problem of measuring the inclination of early-type galaxies is not new, but it is complicated here as the CO is often not aligned with the stars \citep[e.g.][]{Young:2002p943,Schinnerer:2002p981,Young:2008p788,Crocker:2008p946,Crocker:2009p3262,Crocker2010}, and it is the CO inclination that we require. As we have shown, methods using galaxy optical axis ratios are likely to yield large uncertainties and increase the scatter in the TFR, but they do not seem to significantly affect the determination of its slope or zero-point. This is likely because errors in inclination are most problematic when the galaxy is close to face-on, and the number of nearly face-on galaxies in our sample is small. The profile shape selection criterion also actively selects against face-on galaxies, where the velocity width is less than a few channel widths (no matter how extended the molecular gas is).  In many \hi\ TFR papers, galaxies with an inclination less than 45$^{\circ}$ are discarded \citep[e.g. ][]{Tully:1977p2161}, so this is not especially worrying. If we apply this cut in inclination to our sample, we retain within the errors the same Tully-Fisher slope and zero-point, and the RMS scatter decreases. Even with a very harsh cut of 65$^{\circ}$ we retain the same relation. This demonstrates that our results are robust, and the slope and zero-point are minimally affected by the choice of inclination correction.

As hinted above, while the slope and zero-point are robust, the scatter in the relation is affected by the inclination estimation method chosen. Moving from galaxy axial ratios to inclinations derived from dust features decreases the RMS scatter around the best-fit relation by $\approx$0.15 mag, if one removes the one obvious outlier. Our results do reveal a reduction of the total scatter when the inclinations are determined from fitting a thin disc model to interferometric data, despite our limited number statistics. With a larger sample one might expect the scatter to decrease further. In both cases the decrease in scatter is likely due to properly accounting for molecular gas distributions that do not rotate in the same plane as the stars. Unsurprisingly, to obtain the tightest relation one must use measures of inclination (such as unsharp-masked dust images or interferometric imaging of the molecular gas itself) that truly trace the inclination of the gaseous component.

\subsection{Intrinsic scatter of the CO TFR}
The intrinsic scatter of our hybrid TFR is 0.36 mag, broadly consistent with typical values found in dedicated studies of the spiral galaxy TFR in the optical, such as those by \cite{Pizagno:2007p2836} (0.4 mag at $g$- and $z$-band) and \cite{Kannappan:2002p3083} (0.4 mag at $r$-band). Our intrinsic scatter is slightly larger than that found for S0 galaxies by \cite{Williams2010} ($\approx$0.3 mag at $K_{\rm s}$-band). One reason for the increased scatter may be that \cite{Williams2010} considered S0 galaxies only, whereas our sample includes both morphologically classified elliptical and S0 galaxies, that span a wider range in $\lambda_{\rm R}$, the stellar specific angular momentum \citep{Emsellem:2007p1483}. This increased scatter may therefore be due to spanning an increased range in internal dynamics. 

We do not have sufficient number statistics to constrain an elliptical galaxy TFR separately, but the earliest type galaxies (selected by optical morphology) do lie on average to the faster and/or dimmer side of our relations. There is however no systematic behavior within the residuals to the fit as a function of $\lambda_{\rm R}$. 

\subsection{Offset from the spiral TFR}

Exploring changes in the M/L of galaxies as a function of morphological type is one of the main motivations for studying the early-type TFR.
\cite{Williams2010} have highlighted the importance of comparing TFRs for each morphological type derived using the same tracer and methods, in order to minimize systematic differences. In this spirit we use the spiral galaxy line-widths published by \cite{Chung:2009p2772}, measured from on-the-fly mapping of 18 Virgo Cluster members, in order to define a spiral CO TFR. 

\begin{figure}
\begin{center}
\includegraphics[scale=0.525,angle=0,clip,trim=1cm 0cm 0.5cm 1cm]{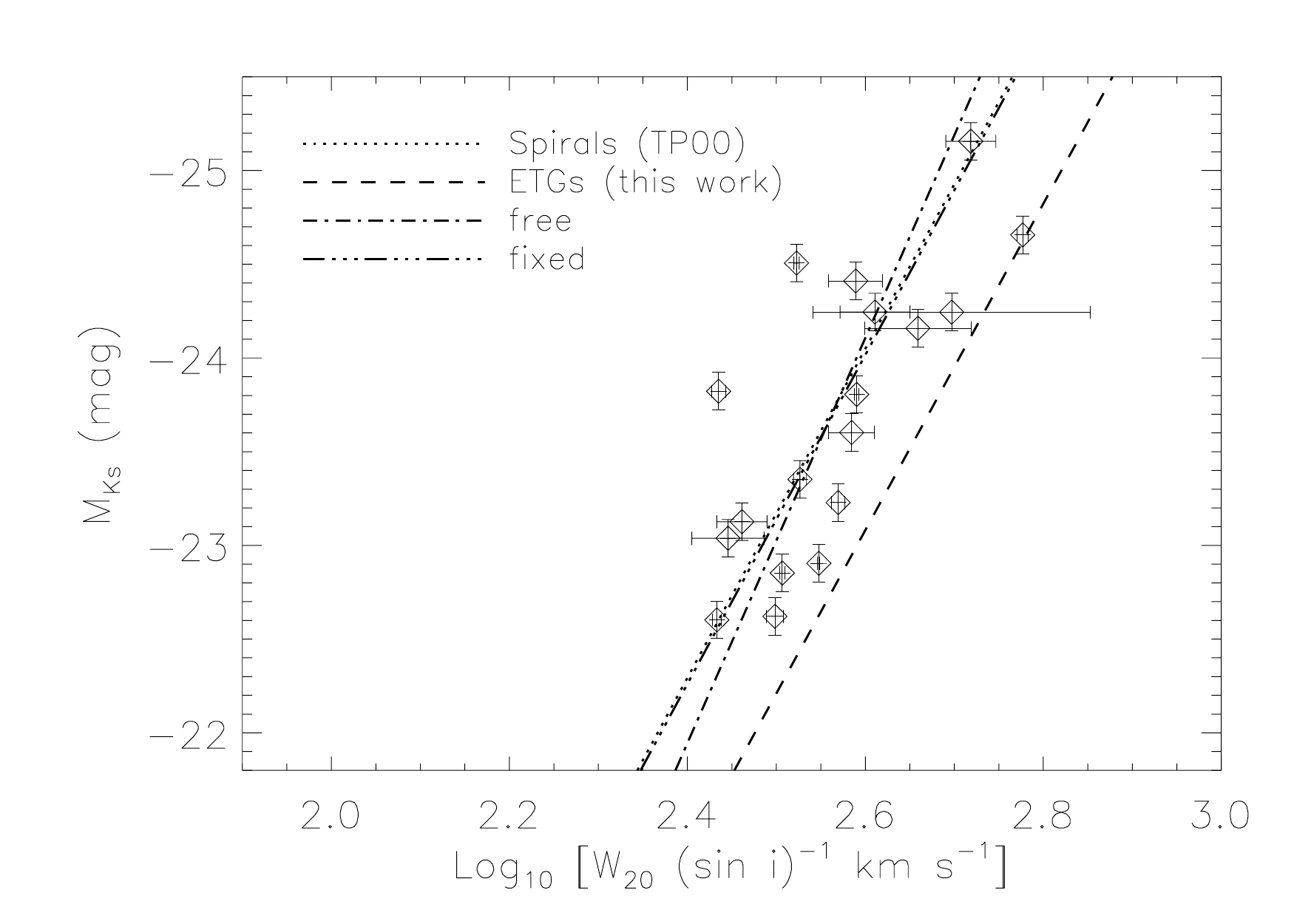}
\end{center}
\caption{\small Spiral galaxy CO TFR built in the same way as that shown in Fig. \ref{tf_rels}, but using W$_{20}$ values and inclinations from the work of \protect \cite{Chung:2009p2772}.}
\label{spiraltfr}
\end{figure}

We construct a Tully-Fisher relation for the \cite{Chung:2009p2772} spirals in the same way as for our early-type sample, using 2MASS $K_{\rm s,total}$ magnitudes and their W$_{20}$ line-widths corrected for inclination. These inclinations were derived from the galaxy axial ratio and the morphological type using the classical Hubble formula from \cite{Hubble:1926p3245}. This formula effectively differs from Equation \ref{galaxial} by making $q_0$ a function of morphological type. For spiral galaxies such as those in the \cite{Chung:2009p2772} sample this correction is small, corresponding to a $q_0$ less than 0.05, and hence this difference is unlikely to systematically effect the comparison of our results. We differ from \cite{Chung:2009p2772} however by not assuming a set distance to the Virgo Cluster, which has significant substructure. Instead we use the distance to each galaxy as determined for the \atlas\ parent sample (Paper I). The resulting TFR is shown in Figure \ref{spiraltfr}. The values used to create this figure are listed in Table \ref{CHUNG_table}.
 
\begin{table}
\caption{\small List of the parameters used to create the spiral TFR in Figure \protect \ref{spiraltfr}.}
\begin{tabular}{c l l c c c} 
\hline
Galaxy & Type &  Distance & m$_{\rm Ks}$& $i_{b/a}$& W$_{20}$ \\
& & (Mpc) & (mag)&(deg) & (\kms)\\
 (1) &(2) &(3) & (4) & (5) & (6)\\
\hline
NGC\,4254  & Sc & 16.5 & 6.93 & 29 $\pm$ 7.3 & 221 $\pm$ 1\\
NGC\,4298  & Sc & 16.5 & 8.47 & 59 $\pm$ 2.4 & 270 $\pm$ 4\\
NGC\,4302  & Sc & 14 & 7.83 & 90 $\pm$ 0.1 & 353 $\pm$ 1\\
NGC\,4303 & Sc & 16.5 & 6.84 & 19 $\pm$ 12. & 162 $\pm$ 2\\
NGC\,4321 & Sc & 15.85 & 6.59 & 38 $\pm$ 5.2 & 239 $\pm$ 2\\
NGC\,4402 & Sc & 16.5 & 8.49 & 80 $\pm$ 0.7 & 267 $\pm$ 3\\
NGC\,4419 & Sa & 13.12 & 7.74 & 82 $\pm$ 0.6 & 318 $\pm$ 2\\
NGC\,4438 & Sb & 16.5 & 7.27 & 87 $\pm$ 0.2 & 272 $\pm$ 5\\
NGC\,4501 & Sbc & 15.3 & 6.27 & 60 $\pm$ 2.3 & 518 $\pm$ 2\\
NGC\,4527 & Sb & 14.05 & 6.93 & 75 $\pm$ 1.1 & 376 $\pm$ 2\\
NGC\,4535 & Sc & 15.75 & 7.38 & 41 $\pm$ 4.7 & 252 $\pm$ 5\\
NGC\,4536 & Sc & 14.09 & 7.52 & 59 $\pm$ 2.4 & 318 $\pm$ 2\\
NGC\,4548 & Sb & 18.71 & 7.12 & 35 $\pm$ 5.8 & 234 $\pm$ 6\\
NGC\,4569 & Sab & 16.5 & 6.58 & 69 $\pm$ 1.6 & 311 $\pm$ 1\\
NGC\,4579 & Sab & 21.3 & 6.49 & 39 $\pm$ 5.0 & 329 $\pm$ 1\\
NGC\,4647 & Sc & 16.5 & 8.05 & 34 $\pm$ 6.0 & 156 $\pm$ 1\\
NGC\,4654 & Sc & 16.5 & 7.74 & 58 $\pm$ 2.5 & 285 $\pm$ 1\\
NGC\,4689 & Sc & 16.5 & 7.96 & 39 $\pm$ 5.0 & 182 $\pm$ 1\\
\hline

\end{tabular}
\label{CHUNG_table}
\textit{Notes:} The galaxy sample listed in Column 1 comes from \protect \cite{Chung:2009p2772}. Column 2 lists the spiral galaxy type following the Hubble scheme, \protect \citep[RC3;][]{deVaucouleurs:1991p2406}. The distance to the galaxy in Column 3 is from the \atlas\ parent sample (Paper I). These distances are drawn preferentially from \cite{Mei:2007p3221} and \cite{Tonry:2001p3222}. Column 4 lists the apparent $K_{\rm{s}}$-band magnitudes, taken from 2MASS. Column 5 is the optical galaxy axis ratio, taken from NED, as described in Section \protect \ref{galaxial}.  The line-widths in Column 6 are taken directly from \protect \cite{Chung:2009p2772}. 
\end{table}

The best-fit spiral TFR with an unconstrained gradient is

\begin{equation}
{\rm M}_{K\rm{s}} =  \rm (-10.8 \pm 2.5)\,[{\rm \log}_{10}\left(\frac{W_{20}}{km\ s^{-1}}\right) - 2.6]-(24.1 \pm 0.2)\,,
\end{equation}
\noindent where the intrinsic scatter about the correlation is 0.68 mag, with a total scatter of 0.72 mag. The TFR obtained by constraining the fit to have the same slope as that of \cite{Tully:2000p2442} is

\begin{equation}
{\rm M}_{K\rm{s}} =  \rm -8.78\,[{\rm \log}_{10}\left(\frac{W_{20}}{km\ s^{-1}}\right) - 2.6]-(24.02 \pm 0.14)\,,
\end{equation}
\noindent with an intrinsic scatter of 0.55 mag and a total scatter of 0.59 mag. 

Within observational errors, the best-fit free and constrained spiral CO TFRs are consistent with the relation of \cite{Tully:2000p2442}. The offset between the unconstrained \cite{Chung:2009p2772} spiral CO TFR and the hybrid early-type CO TFR (discussed in Section \ref{best}) is 0.98 $\pm$ 0.22 mag, consistent within errors with the value found by \cite{Bedregal:2006p2087} and the offset found by \cite{Williams2010} between their S0 TFR and the \cite{Tully:2000p2442} relation. 

This result is nevertheless surprising since \cite{Williams2010} argue that their offset with \cite{Tully:2000p2442} is flawed by systematic effects (the tracers and methods being different), and they prefer the much smaller offset (0.5 $\pm$ 0.14 mag) from their spiral galaxy sample treated identically to the S0s. 

It is not immediately clear how our own result can be reconciled with this, as we use identical tracers and methods as \cite{Chung:2009p2772}. However, we suspect that the different results are due to the mix of morphological types in the spiral samples. \cite{Williams2010} use a spiral sample consisting mostly of Sa-Sb spirals, while the sample of \cite{Chung:2009p2772} includes a larger fraction of later-type galaxies (mostly Sc), which are known to have a larger zero-point \citep[e.g.][]{Roberts:1978p3079,Rubin:1985p3080,Masters:2008p3223,Shen:2009p2679}.
Separating the \cite{Chung:2009p2772} sample more finely into morphological types does not necessarily support this explanation, but the number statistics are very limited.

\subsection{Origin of the offset TFR for ETGs}

In the ETG sample used in this paper, a few galaxies are consistent with the spiral galaxy relation and many lie between the spiral and early-type relations, consistent with the predictions of passive fading models \cite[e.g.][]{Dressler:1980p2456,Dressler:1997p2485}. The fact that molecular gas is present in all these galaxies, however, suggests that many must have ongoing star formation (even if weak), and hence evolutionary models relying on a purely passive evolution  are ruled out. 

The presence of molecular gas mass fractions ranging from $<$0.1 to 5\% in these galaxies perhaps suggests an answer to some of the problems with passive evolution models. \cite{Bedregal:2006p2087} and \cite{Williams2010} both discuss the timescale implied by the ETG-spiral TFR offsets observed, assuming passive fading, and find it uncomfortably short and inconsistent between the $B$ and $K_{\rm s}$ bands. The presence of low levels of residual star formation would prolong the period it takes for galaxies to fade, and would contribute relatively more light at blue wavelengths than in the infrared. The molecular gas mass fraction in these galaxies is however small 
, so it is unclear if such a low mass fraction of young stars could extend the fading timescale sufficiently.

An alternative explanation of the TFR offset is that the morphological (and luminosity) evolution is accompanied by a change in the size of the galaxies, affecting the galaxy rotation rather than its luminosity. \cite{Williams2010} suggest that a systematic contraction as spirals transform into S0s, consistent with the trend with morphological type of the size-luminosity relation \citep{Courteau:2007p3085}, could explain the offset from the spiral TFR. Clearly, a combination of both effects is also possible.

\subsection{High-mass end of the CO TFR}
As discussed in the introduction, one motivation for studying the CO TFR is the ability to probe the full mass range of galaxies, from small disc galaxies to large spheriods, with the same tracer and hence with the same assumptions and systematics. As an example of the power of this approach, we can investigate if our sample shows any evidence of a change in the slope of the TFR at the high-mass end, as reported by some authors \citep[e.g. ][]{Peletier:1993p3082,Verheijen:2001p3081,Noordermeer:2007p3078}. 

Our sample contains eight galaxies above the luminosity threshold suggested in previous works (M$_{K\rm{s}}$ $\leq$-23.75). As can be seen in the bottom-right panel of Figure \ref{tf_rels}, all but one of these systems are on, or to the right of, the best-fit relation for the whole sample. When one compares the best-fit TFR derived for the whole sample to that for just the fainter galaxies (M$_{K\rm{s}}$ $\geq$-23.75), there is a small difference in the gradient and zero-point, in the direction found by other authors. We do not have the number statistics to properly investigate the size of this effect, however, and hence only conclude that our results are consistent with the presence of a different slope at the high-mass end of the TFR.

\subsection{Single-dish vs aperture synthesis}

We have demonstrated in this work that single-dish spectra are sufficient to construct a robust CO Tully-Fisher relation, and that galaxy inclinations can be taken from optical imaging without noticeably biasing its slope or zero-point (Section \ref{discussinc}). However, one must exercise caution, as the small beams of millimeter telescopes can occasionally underestimate the line-widths of galaxies with extended molecular discs. Barring beam dilution effects, one would expect single-dish profiles to perform even better at higher redshifts, where the angular size of the target galaxies decreases. One caveat to this is that as the angular size of galaxies decreases, source confusion may become a problem.   

Naively, interferometric observations should be the best way to obtain an estimate of the true inclination of the molecular gas, and hence reduce the artificial scatter in the TFR. Aperture synthesis observations also tend to have a large field of view, and are thus less likely to face problems with extended gas distributions. Interferometric observations do, however, usually require a larger investment of observing time per source. One must also choose the angular resolution of the data carefully, in order to avoid resolving out the most extended structures, while still having sufficient spatial information to constrain a tilted-ring model.
Both single-dish and interferometric studies of the TFR should thus have a bright future.

\section{Conclusions and Future Prospects}
In this paper we have presented the first CO Tully-Fisher relation for early-type galaxies.
We have shown that CO line widths are a reliable tracer of the circular velocity in fast-rotating ETGs, and relations derived in this way agree well with ETG TFRs derived by other authors.
We showed that our CO TFR relation has 
 a robust slope and zero point, and a comparable scatter to that of TFRs derived with other techniques. The power of the CO Tully-Fisher relation is the ability to perform the same analysis easily for both spiral and early-type galaxies (i.e. identical tracer and method), with purely observable quantities. This technique is therefore particularly suitable for comparing the morphological variations of the zero-point of the Tully-Fisher relation.
 
The results presented in this paper show that early-type galaxies follow a TFR that is offset by nearly one magnitude at $K_{\rm{s}}$-band from that of spiral galaxies. This offset is similar to that found by previous authors. An offset ETG TFR is usually thought to be caused by passive evolution of spiral galaxies into lenticulars, after an abrupt cessation of star formation.
However, the presence of molecular gas in these ETGs suggests that the majority of this sample must have ongoing residual star-formation, such as that revealed in a subsample of these galaxies \citep[e.g.][]{Temi:2009p1440,Shapiro:2010p2932,Crocker2010}. This should increase the timescale to fade by the required $\approx$1 mag. Some models, such as those that involve a change in the size of galaxies as they transform, do not require star formation to completely cease, providing an alternative explanation that is fully consistent with our data. Unfortunately the low molecular gas fractions in our systems make drawing any firm conclusions difficult.

Another area where the CO TFR (for all spectral types) excels is the ability to extend the technique to higher redshifts, as discussed in \cite{Tutui:2001p3401}. \hi\ observations are currently limited to $z\ltsimeq$0.2 \citep[e.g.][]{Catinella:2008p3231}, and stellar kinematics to $z\ltsimeq$2 \citep[e.g.][]{vanDokkum:2009p3229,Cappellari:2009p3228,Cenarro:2009p3230} whereas CO has already been detected in galaxies at redshifts $z>$ 6 \citep[e.g.][]{Walter:2004p3216}. The molecular content of galaxies is also expected to increase with redshift, while the \hi\ content stays approximately constant \citep{Obreschkow:2009p2839,Bauermeister:2009p2677,Daddi:2010p3261,Tacconi:2010p3225}. CO velocity fields can be resolved, and the rotation (and hence the dynamical mass) estimated even out to high redshift \citep[e.g.][]{Walter:2004p3216,Carilli:2010p3263}. 
This raises the exciting prospect of being able to directly track the mass-to-light ratio evolution of galaxies as a function of both redshift and mophological type, using an easily observable tracer (and without requiring dynamical modeling). New optical/IR instrumentation is just starting to allow surveys of galaxy stellar kinematics out to z $\approx$ 2 \citep[e.g.][]{Cappellari:2009p3228,ForsterSchreiber:2009p3246}, and comparison of these samples to line-widths derived with CO is a promising immediate avenue for future studies of the TFR.

\cite{Tutui:2001p3401} showed that constructing a (spiral galaxy) CO TFR is already possible at up to redshifts z$\sim$0.1, but with the next generation of mm-wave telescopes due to enter service over the next few years, we should gain access to CO lines across the redshifted universe for all morphological types. The Redshift Receiver System (RRS) on the Large Millimeter Telescope (LMT), for instance, is optimized for the detection of redshifted transitions of the CO ladder from star-forming galaxies at cosmological distances, and will generate a large database of CO line-widths. If one is able to optically image these galaxies to estimate their inclination and morphology, then this will become an excellent resource for Tully-Fisher analyses.
With the ability to detect spectral line emission from CO in a galaxy like the Milky Way at a z$\sim$3 in less than 24 hours, ALMA will also be in an excellent position to study the CO TFR at high redshift. 
The higher frequency bands that will be available at ALMA also open up the possibility of detecting [CII] emission (rest wavelength 158 $\mu$m), which is brighter than CO in many galaxies in the local universe \citep{Crawford:1985p2859} and possibly enhanced further at high redshifts \citep{Maiolino:2009p2858}. [CII] may therefore become an important tracer of the circular velocity of high-redshift systems.

Although technical aspects will certainly be improved and the interpretation of this relation remains uncertain, we have shown that the ETG CO TFR is a tantalising and promising tool for galaxy evolution.\\

\section*{Acknowledgements}

TAD wishes to thank Gyula Jozsa for help and advice regarding the \tirific\ package and Michael Williams for useful discussions and for providing his TFR fitting code. 
TAD and NS are supported by an STFC Postgraduate Studentship. MC acknowledges support from a STFC Advanced Fellowship (PP/D005574/1) and a Royal Society University Research Fellowship.
This work was supported by the rolling grants `Astrophysics at Oxford' PP/E001114/1 and ST/H002456/1 and visitors grants PPA/V/S/2002/00553, PP/E001564/1 and ST/H504862/1 from the UK Research Councils. RLD acknowledges travel and computer grants from Christ Church, Oxford and support from the Royal Society in the form of a Wolfson Merit Award 502011.K502/jd. RLD also acknowledges the support of the ESO Visitor Programme which funded a 3 month stay in 2010.
RMcD is supported by the Gemini Observatory, which is operated by the Association of Universities for Research in Astronomy, Inc., on behalf of the international Gemini partnership of Argentina, Australia, Brazil, Canada, Chile, the United Kingdom, and the United States of America.
SK acknowledges support from the the Royal Society Joint Projects Grant JP0869822.
TN and MBois acknowledge support from the DFG Cluster of Excellence `Origin and Structure of the Universe'.
MS acknowledges support from a STFC Advanced Fellowship ST/F009186/1.
The authors acknowledge financial support from ESO. 
This paper is partly based on observations carried out with the IRAM 30m telescope. IRAM is supported by INSU/CNRS (France), MPG 
(Germany) and ING (Spain).
Support for CARMA construction was derived from the states of California, Illinois and Maryland, the James S. McDonnell Foundation, the Gordon and Betty Moore Foundation, the Kenneth T. and Eileen L. Norris Foundation, the University of Chicago, the Associates of the California Institute of Technology and the National Science Foundation. Ongoing CARMA development and operations are supported by the National Science Foundation under a cooperative agreement, and by the CARMA partner universities. 
We acknowledge use of
the HYPERLEDA database ({\tt http://leda.univ-lyon1.fr}) and
the NASA/IPAC Extragalactic Database (NED) which is operated
by the Jet Propulsion Laboratory, California Institute of Technology,
under contract with the National Aeronautics and Space Administration.
This publication makes use of data products from the Two Micron All Sky Survey, which is a joint project of the University of Massachusetts and the Infrared Processing and Analysis Center/California Institute of Technology, funded by the National Aeronautics and Space Administration and the National Science Foundation. 

\bsp
\bibliographystyle{mn2e}

\begin{thebibliography}{}

\bibitem[\protect\citeauthoryear{{Abazajian}, {Adelman-McCarthy},
  {Ag{\"u}eros}, {Allam}, {Allende Prieto}, {An}, {Anderson}, {Anderson},
  {Annis}, {Bahcall} \& et al.}{{Abazajian} et~al.}{2009}]{Abazajian:2009p3430}
{Abazajian} K.~N.,  {Adelman-McCarthy} J.~K.,  {Ag{\"u}eros} M.~A.,  {Allam}
  S.~S.,  {Allende Prieto} C.,  {An} D.,  {Anderson} K.~S.~J.,  {Anderson}
  S.~F.,  {Annis} J.,  {Bahcall} N.~A.,    et al. 2009, ApJS, 182, 543

\bibitem[\protect\citeauthoryear{Adelman-McCarthy, Ag{\"u}eros, Allam, Prieto,
  Anderson, Anderson, Annis, Bahcall, Bailer-Jones, Baldry, Barentine, Bassett,
  Becker, Beers, Bell, Berlind \& Bernardi}{Adelman-McCarthy
  et~al.}{2008}]{AdelmanMcCarthy:2008p2834}
Adelman-McCarthy J.~K.,  Ag{\"u}eros M.~A.,  Allam S.~S.,  Prieto C.~A.,
  Anderson K. S.~J.,  Anderson S.~F.,  Annis J.,  Bahcall N.~A.,  Bailer-Jones
  C. A.~L.,  Baldry I.~K.,  Barentine J.~C.,  Bassett B.~A.,  Becker A.~C.,
  Beers T.~C.,  Bell E.~F.,  Berlind A.~A.,    Bernardi M.,  2008, ApJS, 175,
  297

\bibitem[\protect\citeauthoryear{Alatalo, Blitz \& et al}{Alatalo
  et~al.}{2011}]{ngc1266}
Alatalo K.,  Blitz L.,    et al 2011, ApJ

\bibitem[\protect\citeauthoryear{Bauermeister, Blitz \& Ma}{Bauermeister
  et~al.}{2009}]{Bauermeister:2009p2677}
Bauermeister A.,  Blitz L.,    Ma C.-P.,  2009, eprint arXiv, 0909, 3840

\bibitem[\protect\citeauthoryear{Bedregal, Arag{\'o}n-Salamanca \&
  Merrifield}{Bedregal et~al.}{2006}]{Bedregal:2006p2087}
Bedregal A.~G.,  Arag{\'o}n-Salamanca A.,    Merrifield M.~R.,  2006, MNRAS,
  373, 1125

\bibitem[\protect\citeauthoryear{Bell \& de Jong}{Bell \&
  de~Jong}{2001}]{Bell:2001p3237}
Bell E.~F.,  de Jong R.~S.,  2001, ApJ, 550, 212

\bibitem[\protect\citeauthoryear{Bertola, Cinzano, Corsini, Rix \&
  Zeilinger}{Bertola et~al.}{1995}]{Bertola:1995p3239}
Bertola F.,  Cinzano P.,  Corsini E.~M.,  Rix H.-W.,    Zeilinger W.~W.,  1995,
  ApJ Letters v.448, 448, L13

\bibitem[\protect\citeauthoryear{Bock, Bolatto, Hawkins, Kemball, Lamb,
  Plambeck, Pound, Scott, Woody \& Wright}{Bock et~al.}{2006}]{Bock:2006p2806}
Bock D. C.-J.,  Bolatto A.~D.,  Hawkins D.~W.,  Kemball A.~J.,  Lamb J.~W.,
  Plambeck R.~L.,  Pound M.~W.,  Scott S.~L.,  Woody D.~P.,    Wright M. C.~H.,
   2006, Ground-based and Airborne Telescopes. Proc. SPIE, 6267, 13

\bibitem[\protect\citeauthoryear{Buta}{Buta}{1986}]{Buta:1986p3227}
Buta R.,  1986, ApJ Supplement v.61, 61, 631

\bibitem[\protect\citeauthoryear{Cappellari}{Cappellari}{2008}]{Cappellari:200%
8p2773}
Cappellari M.,  2008, MNRAS, 390, 71

\bibitem[\protect\citeauthoryear{Cappellari, Bacon, Bureau, Damen, Davies, de
  Zeeuw, Emsellem, Falc{\'o}n-Barroso, Krajnovi{\'c}, Kuntschner, McDermid,
  Peletier, Sarzi, van~den Bosch \& van~de Ven}{Cappellari
  et~al.}{2006}]{Cappellari:2006p1498}
Cappellari M.,  Bacon R.,  Bureau M.,  Damen M.~C.,  Davies R.~L.,  de Zeeuw
  P.~T.,  Emsellem E.,  Falc{\'o}n-Barroso J.,  Krajnovi{\'c} D.,  Kuntschner
  H.,  McDermid R.~M.,  Peletier R.~F.,  Sarzi M.,  van~den Bosch R. C.~E.,
  van~de Ven G.,  2006, MNRAS, 366, 1126

\bibitem[\protect\citeauthoryear{Cappellari, di Serego~Alighieri, Cimatti,
  Daddi, Renzini, Kurk, Cassata, Dickinson, Franceschini, Mignoli, Pozzetti,
  Rodighiero, Rosati \& Zamorani}{Cappellari
  et~al.}{2009}]{Cappellari:2009p3228}
Cappellari M.,  di Serego~Alighieri S.,  Cimatti A.,  Daddi E.,  Renzini A.,
  Kurk J.~D.,  Cassata P.,  Dickinson M.,  Franceschini A.,  Mignoli M.,
  Pozzetti L.,  Rodighiero G.,  Rosati P.,    Zamorani G.,  2009, ApJ Letters,
  704, L34

\bibitem[\protect\citeauthoryear{Cappellari, Emsellem, Krajnovic, McDermid,
  Scott, Kleijn, Young, Alatalo, Bacon, Blitz, Bois, Bournaud, Bureau, Davies,
  Davis, de Zeeuw, Duc, Khochfar, Kuntschner \& et al.}{Cappellari
  et~al.}{2011}]{Cap2010}
Cappellari M.,  Emsellem E.,  Krajnovic D.,  McDermid R.~M.,  Scott N.,  Kleijn
  G. A.~V.,  Young L.~M.,  Alatalo K.,  Bacon R.,  Blitz L.,  Bois M.,
  Bournaud F.,  Bureau M.,  Davies R.~L.,  Davis T.~A.,  de Zeeuw P.~T.,  Duc
  P.-A.,  Khochfar S.,  Kuntschner H.,    et al. 2011, eprint arXiv (Paper I),
  1012, 1551

\bibitem[\protect\citeauthoryear{Cappellari, Scott, Alatalo, Blitz, Bois,
  Bournaud, Bureau, Davies, Davis, de Zeeuw, Emsellem, Falcon-Barroso,
  Khochfar, Krajnovic, Kuntschner \& et. al.}{Cappellari
  et~al.}{2010}]{Cappellari:2010p3429}
Cappellari M.,  Scott N.,  Alatalo K.,  Blitz L.,  Bois M.,  Bournaud F.,
  Bureau M.,  Davies R.~L.,  Davis T.~A.,  de Zeeuw P.~T.,  Emsellem E.,
  Falcon-Barroso J.,  Khochfar S.,  Krajnovic D.,  Kuntschner H.,    et. al.
  2010, Highlights of Astronomy, 15, 81

\bibitem[\protect\citeauthoryear{Carilli, Daddi, Riechers, Walter, Weiss,
  Dannerbauer, Morrison, Wagg, Dav{\'e}, Elbaz, Stern, Dickinson, Krips \&
  Aravena}{Carilli et~al.}{2010}]{Carilli:2010p3263}
Carilli C.~L.,  Daddi E.,  Riechers D.,  Walter F.,  Weiss A.,  Dannerbauer H.,
   Morrison G.~E.,  Wagg J.,  Dav{\'e} R.,  Elbaz D.,  Stern D.,  Dickinson M.,
   Krips M.,    Aravena M.,  2010, ApJ, 714, 1407

\bibitem[\protect\citeauthoryear{Catinella}{Catinella}{2008}]{Catinella:2008p3%
231}
Catinella B.,  2008, The evolution of galaxies through the neutral hydrogen
  window. AIP Conference Proceedings, 1035, 186

\bibitem[\protect\citeauthoryear{Cenarro \& Trujillo}{Cenarro \&
  Trujillo}{2009}]{Cenarro:2009p3230}
Cenarro A.~J.,  Trujillo I.,  2009, ApJ Letters, 696, L43

\bibitem[\protect\citeauthoryear{Christodoulou, Tohline \&
  Steiman-Cameron}{Christodoulou et~al.}{1988}]{Christodoulou:1988p3010}
Christodoulou D.~M.,  Tohline J.~E.,    Steiman-Cameron T.~Y.,  1988, AJ, 96,
  1307

\bibitem[\protect\citeauthoryear{Chung, Rhee, Kim, Yun, Heyer \& Young}{Chung
  et~al.}{2009}]{Chung:2009p2772}
Chung E.~J.,  Rhee M.-H.,  Kim H.,  Yun M.~S.,  Heyer M.,    Young J.~S.,
  2009, ApJS, 184, 199

\bibitem[\protect\citeauthoryear{Courteau, Dutton, van~den Bosch, MacArthur,
  Dekel, McIntosh \& Dale}{Courteau et~al.}{2007}]{Courteau:2007p3085}
Courteau S.,  Dutton A.~A.,  van~den Bosch F.~C.,  MacArthur L.~A.,  Dekel A.,
  McIntosh D.~H.,    Dale D.~A.,  2007, ApJ, 671, 203

\bibitem[\protect\citeauthoryear{Crawford, Genzel, Townes \& Watson}{Crawford
  et~al.}{1985}]{Crawford:1985p2859}
Crawford M.~K.,  Genzel R.,  Townes C.~H.,    Watson D.~M.,  1985, ApJ, 291,
  755

\bibitem[\protect\citeauthoryear{Crocker, Bureau, Young \& Combes}{Crocker
  et~al.}{2008}]{Crocker:2008p946}
Crocker A.~F.,  Bureau M.,  Young L.~M.,    Combes F.,  2008, MNRAS, 386, 1811

\bibitem[\protect\citeauthoryear{Crocker, Bureau, Young \& Combes}{Crocker
  et~al.}{2010}]{Crocker2010}
Crocker A.~F.,  Bureau M.,  Young L.~M.,    Combes F.,  2010, MNRAS

\bibitem[\protect\citeauthoryear{Crocker, Jeong, Komugi, Combes, Bureau, Young
  \& Yi}{Crocker et~al.}{2009}]{Crocker:2009p3262}
Crocker A.~F.,  Jeong H.,  Komugi S.,  Combes F.,  Bureau M.,  Young L.~M.,
  Yi S.,  2009, MNRAS, 393, 1255

\bibitem[\protect\citeauthoryear{Daddi, Bournaud, Walter, Dannerbauer, Carilli,
  Dickinson, Elbaz, Morrison, Riechers, Onodera, Salmi, Krips \& Stern}{Daddi
  et~al.}{2010}]{Daddi:2010p3261}
Daddi E.,  Bournaud F.,  Walter F.,  Dannerbauer H.,  Carilli C.~L.,  Dickinson
  M.,  Elbaz D.,  Morrison G.~E.,  Riechers D.,  Onodera M.,  Salmi F.,  Krips
  M.,    Stern D.,  2010, ApJ, 713, 686

\bibitem[\protect\citeauthoryear{de Vaucouleurs, de Vaucouleurs, Corwin, Buta,
  Paturel \& Fouque}{de~Vaucouleurs et~al.}{1991}]{deVaucouleurs:1991p2406}
de Vaucouleurs G.,  de Vaucouleurs A.,  Corwin H.~G.,  Buta R.~J.,  Paturel G.,
     Fouque P.,  1991, Volume 1-3

\bibitem[\protect\citeauthoryear{de Zeeuw, Bureau, Emsellem, Bacon, Carollo,
  Copin, Davies, Kuntschner, Miller, Monnet, Peletier \& Verolme}{de~Zeeuw
  et~al.}{2002}]{deZeeuw:2002p1496}
de Zeeuw P.~T.,  Bureau M.,  Emsellem E.,  Bacon R.,  Carollo C.~M.,  Copin Y.,
   Davies R.~L.,  Kuntschner H.,  Miller B.~W.,  Monnet G.,  Peletier R.~F.,
  Verolme E.~K.,  2002, MNRAS, 329, 513

\bibitem[\protect\citeauthoryear{Dickey \& Kazes}{Dickey \&
  Kazes}{1992}]{Dickey:1992p2418}
Dickey J.~M.,  Kazes I.,  1992, ApJ, 393, 530

\bibitem[\protect\citeauthoryear{Dressler}{Dressler}{1980}]{Dressler:1980p2456}
Dressler A.,  1980, ApJ, 236, 351

\bibitem[\protect\citeauthoryear{Dressler, Oemler, Couch, Smail, Ellis, Barger,
  Butcher, Poggianti \& Sharples}{Dressler et~al.}{1997}]{Dressler:1997p2485}
Dressler A.,  Oemler A.,  Couch W.~J.,  Smail I.,  Ellis R.~S.,  Barger A.,
  Butcher H.,  Poggianti B.~M.,    Sharples R.~M.,  1997, ApJ v.490, 490, 577

\bibitem[\protect\citeauthoryear{Dutton, Conroy, van~den Bosch, Prada \&
  More}{Dutton et~al.}{2010}]{Dutton2010}
Dutton A.~A.,  Conroy C.,  van~den Bosch F.~C.,  Prada F.,    More S.,  2010,
  MNRAS, p.~923

\bibitem[\protect\citeauthoryear{Emsellem, Cappellari \& et al}{Emsellem
  et~al.}{2010}]{Emsellem2010}
Emsellem E.,  Cappellari M.,    et al 2010, MNRAS (Paper III)

\bibitem[\protect\citeauthoryear{Emsellem, Cappellari, Krajnovi{\'c}, van~de
  Ven, Bacon, Bureau, Davies, de Zeeuw, Falc{\'o}n-Barroso, Kuntschner,
  McDermid, Peletier \& Sarzi}{Emsellem et~al.}{2007}]{Emsellem:2007p1483}
Emsellem E.,  Cappellari M.,  Krajnovi{\'c} D.,  van~de Ven G.,  Bacon R.,
  Bureau M.,  Davies R.~L.,  de Zeeuw P.~T.,  Falc{\'o}n-Barroso J.,
  Kuntschner H.,  McDermid R.,  Peletier R.~F.,    Sarzi M.,  2007, MNRAS, 379,
  401

\bibitem[\protect\citeauthoryear{Emsellem, Cappellari, Peletier, McDermid,
  Bacon, Bureau, Copin, Davies, Krajnovi{\'c}, Kuntschner, Miller \& de
  Zeeuw}{Emsellem et~al.}{2004}]{Emsellem:2004p1497}
Emsellem E.,  Cappellari M.,  Peletier R.~F.,  McDermid R.~M.,  Bacon R.,
  Bureau M.,  Copin Y.,  Davies R.~L.,  Krajnovi{\'c} D.,  Kuntschner H.,
  Miller B.~W.,    de Zeeuw P.~T.,  2004, MNRAS, 352, 721

\bibitem[\protect\citeauthoryear{{Emsellem}, {Monnet} \& {Bacon}}{{Emsellem}
  et~al.}{1994}]{Emsellem:1994p723}
{Emsellem} E.,  {Monnet} G.,    {Bacon} R.,  1994, A\&A, 285, 723

\bibitem[\protect\citeauthoryear{Faber \& Jackson}{Faber \&
  Jackson}{1976}]{Faber:1976p3242}
Faber S.~M.,  Jackson R.~E.,  1976, ApJ, 204, 668

\bibitem[\protect\citeauthoryear{F{\"o}rster-Schreiber, Genzel, Bouch{\'e},
  Cresci, Davies, Buschkamp, Shapiro \& Tacconi}{F{\"o}rster-Schreiber
  et~al.}{2009}]{ForsterSchreiber:2009p3246}
F{\"o}rster-Schreiber N.~M.,  Genzel R.,  Bouch{\'e} N.,  Cresci G.,  Davies
  R.,  Buschkamp P.,  Shapiro K.,    Tacconi L.~J.,  2009, ApJ, 706, 1364

\bibitem[\protect\citeauthoryear{Gavazzi}{Gavazzi}{1993}]{Gavazzi:1993p2448}
Gavazzi G.,  1993, ApJ v.419, 419, 469

\bibitem[\protect\citeauthoryear{Gavazzi, Treu, Rhodes, Koopmans, Bolton,
  Burles, Massey \& Moustakas}{Gavazzi et~al.}{2007}]{Gavazzi:2007p2831}
Gavazzi R.,  Treu T.,  Rhodes J.~D.,  Koopmans L. V.~E.,  Bolton A.~S.,  Burles
  S.,  Massey R.~J.,    Moustakas L.~A.,  2007, ApJ, 667, 176

\bibitem[\protect\citeauthoryear{Gerhard, Kronawitter, Saglia \&
  Bender}{Gerhard et~al.}{2001}]{Gerhard:2001p3046}
Gerhard O.,  Kronawitter A.,  Saglia R.~P.,    Bender R.,  2001, AJ, 121, 1936

\bibitem[\protect\citeauthoryear{Ho}{Ho}{2007}]{Ho:2007p2426}
Ho L.~C.,  2007, ApJ, 669, 821

\bibitem[\protect\citeauthoryear{Hubble}{Hubble}{1926}]{Hubble:1926p3245}
Hubble E.~P.,  1926, ApJ, 64, 321

\bibitem[\protect\citeauthoryear{Jarrett, Chester, Cutri, Schneider, Skrutskie
  \& Huchra}{Jarrett et~al.}{2000}]{Jarrett:2000p2407}
Jarrett T.~H.,  Chester T.,  Cutri R.,  Schneider S.,  Skrutskie M.,    Huchra
  J.~P.,  2000, AJ, 119, 2498

\bibitem[\protect\citeauthoryear{J{\'o}zsa, Kenn, Klein \& Oosterloo}{J{\'o}zsa
  et~al.}{2007}]{Jozsa:2007p2673}
J{\'o}zsa G. I.~G.,  Kenn F.,  Klein U.,    Oosterloo T.~A.,  2007, A\&A, 468,
  731

\bibitem[\protect\citeauthoryear{J{\'o}zsa, Oosterloo, Morganti, Klein \&
  Erben}{J{\'o}zsa et~al.}{2009}]{Jozsa:2009p3232}
J{\'o}zsa G. I.~G.,  Oosterloo T.~A.,  Morganti R.,  Klein U.,    Erben T.,
  2009, A\&A, 494, 489

\bibitem[\protect\citeauthoryear{Kannappan, Fabricant \& Franx}{Kannappan
  et~al.}{2002}]{Kannappan:2002p3083}
Kannappan S.~J.,  Fabricant D.~G.,    Franx M.,  2002, AJ, 123, 2358

\bibitem[\protect\citeauthoryear{Kent}{Kent}{1987}]{Kent:1987p2832}
Kent S.~M.,  1987, AJ, 93, 816

\bibitem[\protect\citeauthoryear{Koribalski, Dahlem, Mebold \&
  Brinks}{Koribalski et~al.}{1993}]{Koribalski:1993p3012}
Koribalski B.,  Dahlem M.,  Mebold U.,    Brinks E.,  1993, A\&A, 268, 14

\bibitem[\protect\citeauthoryear{Krajnovi{\'c}, Emsellem \& et
  al}{Krajnovi{\'c} et~al.}{2010}]{Kraj2010}
Krajnovi{\'c} D.,  Emsellem E.,    et al 2010, MNRAS (Paper II)

\bibitem[\protect\citeauthoryear{Kron}{Kron}{1980}]{Kron:1980p3008}
Kron R.~G.,  1980, ApJS, 43, 305

\bibitem[\protect\citeauthoryear{Lavezzi}{Lavezzi}{1997}]{Lavezzi:1997p2421}
Lavezzi T.~E.,  1997, PhD Thesis, University of Minnesota, p.~5

\bibitem[\protect\citeauthoryear{Lavezzi \& Dickey}{Lavezzi \&
  Dickey}{1997}]{Lavezzi:1997p2868}
Lavezzi T.~E.,  Dickey J.~M.,  1997, Astronomical Journal v.114, 114, 2437

\bibitem[\protect\citeauthoryear{Lavezzi \& Dickey}{Lavezzi \&
  Dickey}{1998}]{Lavezzi:1998p2411}
Lavezzi T.~E.,  Dickey J.~M.,  1998, AJ, 116, 2672

\bibitem[\protect\citeauthoryear{Magorrian \& Ballantyne}{Magorrian \&
  Ballantyne}{2001}]{Magorrian:2001p3045}
Magorrian J.,  Ballantyne D.,  2001, MNRAS, 322, 702

\bibitem[\protect\citeauthoryear{Maiolino, Caselli, Nagao, Walmsley, Breuck \&
  Meneghetti}{Maiolino et~al.}{2009}]{Maiolino:2009p2858}
Maiolino R.,  Caselli P.,  Nagao T.,  Walmsley M.,  Breuck C.~D.,    Meneghetti
  M.,  2009, A\&A, 500, L1

\bibitem[\protect\citeauthoryear{Markwardt}{Markwardt}{2009}]{Markwardt:2009p2%
588}
Markwardt C.~B.,  2009, Astronomical Data Analysis Software and Systems XVIII
  ASP Conference Series, 411, 251

\bibitem[\protect\citeauthoryear{Masters, Springob \& Huchra}{Masters
  et~al.}{2008}]{Masters:2008p3223}
Masters K.~L.,  Springob C.~M.,    Huchra J.~P.,  2008, AJ, 135, 1738

\bibitem[\protect\citeauthoryear{Mei, Blakeslee, C{\^o}t{\'e}, Tonry, West,
  Ferrarese, Jord{\'a}n, Peng, Anthony \& Merritt}{Mei
  et~al.}{2007}]{Mei:2007p3221}
Mei S.,  Blakeslee J.~P.,  C{\^o}t{\'e} P.,  Tonry J.~L.,  West M.~J.,
  Ferrarese L.,  Jord{\'a}n A.,  Peng E.~W.,  Anthony A.,    Merritt D.,  2007,
  ApJ, 655, 144

\bibitem[\protect\citeauthoryear{Morganti, de Zeeuw, Oosterloo, McDermid,
  Krajnovi{\'c}, Cappellari, Kenn, Weijmans \& Sarzi}{Morganti
  et~al.}{2006}]{Morganti:2006p1934}
Morganti R.,  de Zeeuw P.~T.,  Oosterloo T.~A.,  McDermid R.~M.,  Krajnovi{\'c}
  D.,  Cappellari M.,  Kenn F.,  Weijmans A.,    Sarzi M.,  2006, MNRAS, 371,
  157

\bibitem[\protect\citeauthoryear{Neistein, Maoz, Rix \& Tonry}{Neistein
  et~al.}{1999}]{Neistein:1999p3044}
Neistein E.,  Maoz D.,  Rix H.-W.,    Tonry J.~L.,  1999, AJ, 117, 2666

\bibitem[\protect\citeauthoryear{Nilson}{Nilson}{1973}]{Nilson:1973p3009}
Nilson P.,  1973, Acta Universitatis Upsaliensis. Nova Acta Regiae Societatis
  Scientiarum Upsaliensis - Uppsala Astronomiska Observatoriums Annaler

\bibitem[\protect\citeauthoryear{Noordermeer \& Verheijen}{Noordermeer \&
  Verheijen}{2007}]{Noordermeer:2007p3078}
Noordermeer E.,  Verheijen M. A.~W.,  2007, MNRAS, 381, 1463

\bibitem[\protect\citeauthoryear{Obreschkow \& Rawlings}{Obreschkow \&
  Rawlings}{2009}]{Obreschkow:2009p2839}
Obreschkow D.,  Rawlings S.,  2009, ApJ Letters, 696, L129

\bibitem[\protect\citeauthoryear{Okuda, Kohno, Iguchi \& Nakanishi}{Okuda
  et~al.}{2005}]{Okuda:2005p3259}
Okuda T.,  Kohno K.,  Iguchi S.,    Nakanishi K.,  2005, ApJ, 620, 673

\bibitem[\protect\citeauthoryear{Oosterloo, Morganti, Crocker, Juette,
  Cappellari, de Zeeuw, Krajnovic, McDermid, Kuntschner, Sarzi \&
  Weijmans}{Oosterloo et~al.}{2010}]{Oosterloo:2010p3287}
Oosterloo T.,  Morganti R.,  Crocker A.,  Juette E.,  Cappellari M.,  de Zeeuw
  T.,  Krajnovic D.,  McDermid R.,  Kuntschner H.,  Sarzi M.,    Weijmans
  A.-M.,  2010, eprint arXiv, 1007, 2059

\bibitem[\protect\citeauthoryear{Oosterloo, Morganti, Sadler, van~der Hulst \&
  Serra}{Oosterloo et~al.}{2007}]{Oosterloo:2007p3288}
Oosterloo T.~A.,  Morganti R.,  Sadler E.~M.,  van~der Hulst T.,    Serra P.,
  2007, A\&A, 465, 787

\bibitem[\protect\citeauthoryear{{Paturel}, {Petit}, {Prugniel}, {Theureau},
  {Rousseau}, {Brouty}, {Dubois} \& {Cambr{\'e}sy}}{{Paturel}
  et~al.}{2003}]{Paturel:2003p3431}
{Paturel} G.,  {Petit} C.,  {Prugniel} P.,  {Theureau} G.,  {Rousseau} J.,
  {Brouty} M.,  {Dubois} P.,    {Cambr{\'e}sy} L.,  2003, A\&A, 412, 45

\bibitem[\protect\citeauthoryear{Peletier \& Willner}{Peletier \&
  Willner}{1993}]{Peletier:1993p3082}
Peletier R.~F.,  Willner S.~P.,  1993, ApJ v.418, 418, 626

\bibitem[\protect\citeauthoryear{Phillipps}{Phillipps}{1989}]{Phillipps:1989p3%
235}
Phillipps S.,  1989, A\&A, 211, 259

\bibitem[\protect\citeauthoryear{Pizagno, Prada, Weinberg, Rix, Pogge, Grebel,
  Harbeck, Blanton, Brinkmann \& Gunn}{Pizagno
  et~al.}{2007}]{Pizagno:2007p2836}
Pizagno J.,  Prada F.,  Weinberg D.~H.,  Rix H.-W.,  Pogge R.~W.,  Grebel
  E.~K.,  Harbeck D.,  Blanton M.,  Brinkmann J.,    Gunn J.~E.,  2007, AJ,
  134, 945

\bibitem[\protect\citeauthoryear{Rijcke, Zeilinger, Hau, Prugniel \&
  Dejonghe}{Rijcke et~al.}{2007}]{DeRijcke:2007p2492}
Rijcke S.~D.,  Zeilinger W.~W.,  Hau G. K.~T.,  Prugniel P.,    Dejonghe H.,
  2007, ApJ, 659, 1172

\bibitem[\protect\citeauthoryear{Roberts}{Roberts}{1978}]{Roberts:1978p3079}
Roberts M.~S.,  1978, Astronomical Journal, 83, 1026

\bibitem[\protect\citeauthoryear{Rogstad, Lockhart \& Wright}{Rogstad
  et~al.}{1974}]{Rogstad:1974p2774}
Rogstad D.~H.,  Lockhart I.~A.,    Wright M. C.~H.,  1974, ApJ, 193, 309

\bibitem[\protect\citeauthoryear{Rubin, Burstein, Ford \& Thonnard}{Rubin
  et~al.}{1985}]{Rubin:1985p3080}
Rubin V.~C.,  Burstein D.,  Ford W.~K.,    Thonnard N.,  1985, ApJ, 289, 81

\bibitem[\protect\citeauthoryear{Sancisi}{Sancisi}{2004}]{Sancisi:2004p2830}
Sancisi R.,  2004, International Astronomical Union Symposium no. 220, 220, 233

\bibitem[\protect\citeauthoryear{Sault, Teuben \& Wright}{Sault
  et~al.}{1995}]{Sault:1995p2768}
Sault R.~J.,  Teuben P.~J.,    Wright M. C.~H.,  1995, Astronomical Data
  Analysis Software and Systems IV, 77, 433

\bibitem[\protect\citeauthoryear{Schinnerer \& Scoville}{Schinnerer \&
  Scoville}{2002}]{Schinnerer:2002p981}
Schinnerer E.,  Scoville N.,  2002, ApJ, 577, L103

\bibitem[\protect\citeauthoryear{Schoeniger \& Sofue}{Schoeniger \&
  Sofue}{1997}]{Schoeniger:1997p2424}
Schoeniger F.,  Sofue Y.,  1997, A\&A, 323, 14

\bibitem[\protect\citeauthoryear{Schoniger \& Sofue}{Schoniger \&
  Sofue}{1994}]{Schoniger:1994p2416}
Schoniger F.,  Sofue Y.,  1994, A\&A, 283, 21

\bibitem[\protect\citeauthoryear{Scott, Cappellari, Davies, Bacon, de Zeeuw,
  Emsellem, Falc{\'o}n-Barroso, Krajnovi{\'c}, Kuntschner, McDermid, Peletier,
  Pipino, Sarzi, van~den Bosch, van~de Ven \& van Scherpenzeel}{Scott
  et~al.}{2009}]{Scott:2009p3218}
Scott N.,  Cappellari M.,  Davies R.~L.,  Bacon R.,  de Zeeuw P.~T.,  Emsellem
  E.,  Falc{\'o}n-Barroso J.,  Krajnovi{\'c} D.,  Kuntschner H.,  McDermid
  R.~M.,  Peletier R.~F.,  Pipino A.,  Sarzi M.,  van~den Bosch R. C.~E.,
  van~de Ven G.,    van Scherpenzeel E.,  2009, MNRAS, 398, 1835

\bibitem[\protect\citeauthoryear{Shapiro, Falc{\'o}n-Barroso, van~de Ven, de
  Zeeuw, Sarzi, Bacon, Bolatto, Cappellari, Croton, Davies, Emsellem, Fakhouri,
  Krajnovi{\'c}, Kuntschner, McDermid, Peletier, van~den Bosch \& van~der
  Wolk}{Shapiro et~al.}{2010}]{Shapiro:2010p2932}
Shapiro K.~L.,  Falc{\'o}n-Barroso J.,  van~de Ven G.,  de Zeeuw P.~T.,  Sarzi
  M.,  Bacon R.,  Bolatto A.,  Cappellari M.,  Croton D.,  Davies R.~L.,
  Emsellem E.,  Fakhouri O.,  Krajnovi{\'c} D.,  Kuntschner H.,  McDermid
  R.~M.,  Peletier R.~F.,  van~den Bosch R. C.~E.,    van~der Wolk G.,  2010,
  MNRAS, 402, 2140

\bibitem[\protect\citeauthoryear{Shen, Wang, Chang, Shao, Hou \& Shu}{Shen
  et~al.}{2009}]{Shen:2009p2679}
Shen S.,  Wang C.,  Chang R.,  Shao Z.,  Hou J.,    Shu C.,  2009, ApJ, 705,
  1496

\bibitem[\protect\citeauthoryear{Skrutskie, Cutri, Stiening, Weinberg,
  Schneider, Carpenter, Beichman, Capps, Chester, Elias, Huchra, Liebert,
  Lonsdale \& Monet}{Skrutskie et~al.}{2006}]{Skrutskie:2006p2829}
Skrutskie M.~F.,  Cutri R.~M.,  Stiening R.,  Weinberg M.~D.,  Schneider S.,
  Carpenter J.~M.,  Beichman C.,  Capps R.,  Chester T.,  Elias J.,  Huchra J.,
   Liebert J.,  Lonsdale C.,    Monet D.~G.,  2006, AJ, 131, 1163

\bibitem[\protect\citeauthoryear{Sprayberry, Bernstein, Impey \&
  Bothun}{Sprayberry et~al.}{1995}]{Sprayberry:1995p3236}
Sprayberry D.,  Bernstein G.~M.,  Impey C.~D.,    Bothun G.~D.,  1995, ApJ,
  438, 72

\bibitem[\protect\citeauthoryear{Steiman-Cameron, Kormendy \&
  Durisen}{Steiman-Cameron et~al.}{1992}]{SteimanCameron:1992p2507}
Steiman-Cameron T.~Y.,  Kormendy J.,    Durisen R.~H.,  1992, AJ, 104, 1339

\bibitem[\protect\citeauthoryear{Tacconi, Genzel, Neri, Cox, Cooper, Shapiro,
  Bolatto, Bouch{\'e}, Bournaud, Burkert, Combes, Comerford, Davis, Schreiber,
  Garcia-Burillo, Gracia-Carpio, Lutz, Naab, Omont, Shapley, Sternberg \&
  Weiner}{Tacconi et~al.}{2010}]{Tacconi:2010p3225}
Tacconi L.~J.,  Genzel R.,  Neri R.,  Cox P.,  Cooper M.~C.,  Shapiro K.,
  Bolatto A.,  Bouch{\'e} N.,  Bournaud F.,  Burkert A.,  Combes F.,  Comerford
  J.,  Davis M.,  Schreiber N. M.~F.,  Garcia-Burillo S.,  Gracia-Carpio J.,
  Lutz D.,  Naab T.,  Omont A.,  Shapley A.,  Sternberg A.,    Weiner B.,
  2010, Nature, 463, 781

\bibitem[\protect\citeauthoryear{Temi, Brighenti \& Mathews}{Temi
  et~al.}{2009}]{Temi:2009p1440}
Temi P.,  Brighenti F.,    Mathews W.~G.,  2009, ApJ, 695, 1

\bibitem[\protect\citeauthoryear{Tonry, Dressler, Blakeslee, Ajhar, Fletcher,
  Luppino, Metzger \& Moore}{Tonry et~al.}{2001}]{Tonry:2001p3222}
Tonry J.~L.,  Dressler A.,  Blakeslee J.~P.,  Ajhar E.~A.,  Fletcher A.~B.,
  Luppino G.~A.,  Metzger M.~R.,    Moore C.~B.,  2001, ApJ, 546, 681

\bibitem[\protect\citeauthoryear{Toribio \& Solanes}{Toribio \&
  Solanes}{2009}]{Toribio:2009p2676}
Toribio M.~C.,  Solanes J.~M.,  2009, AJ, 138, 1957

\bibitem[\protect\citeauthoryear{Tully \& Fisher}{Tully \&
  Fisher}{1977}]{Tully:1977p2161}
Tully R.~B.,  Fisher J.~R.,  1977, A\&A, 54, 661

\bibitem[\protect\citeauthoryear{Tully \& Fouque}{Tully \&
  Fouque}{1985}]{Tully:1985p2513}
Tully R.~B.,  Fouque P.,  1985, ApJS, 58, 67

\bibitem[\protect\citeauthoryear{Tully \& Pierce}{Tully \&
  Pierce}{2000}]{Tully:2000p2442}
Tully R.~B.,  Pierce M.~J.,  2000, ApJ, 533, 744

\bibitem[\protect\citeauthoryear{Tutui \& Sofue}{Tutui \&
  Sofue}{1997}]{Tutui:1997p2423}
Tutui Y.,  Sofue Y.,  1997, A\&A, 326, 915

\bibitem[\protect\citeauthoryear{Tutui \& Sofue}{Tutui \&
  Sofue}{1999}]{Tutui:1999p2518}
Tutui Y.,  Sofue Y.,  1999, A\&A, 351, 467

\bibitem[\protect\citeauthoryear{Tutui, Sofue, Honma, Ichikawa \&
  Wakamatsu}{Tutui et~al.}{2001}]{Tutui:2001p3401}
Tutui Y.,  Sofue Y.,  Honma M.,  Ichikawa T.,    Wakamatsu K.-I.,  2001, PASJ,
  53, 701

\bibitem[\protect\citeauthoryear{van Dokkum, Kriek \& Franx}{van Dokkum
  et~al.}{2009}]{vanDokkum:2009p3229}
van Dokkum P.~G.,  Kriek M.,    Franx M.,  2009, Nature, 460, 717

\bibitem[\protect\citeauthoryear{Verheijen}{Verheijen}{2001}]{Verheijen:2001p3%
081}
Verheijen M. A.~W.,  2001, ApJ, 563, 694

\bibitem[\protect\citeauthoryear{Walter, Carilli, Bertoldi, Menten, Cox, Lo,
  Fan \& Strauss}{Walter et~al.}{2004}]{Walter:2004p3216}
Walter F.,  Carilli C.,  Bertoldi F.,  Menten K.,  Cox P.,  Lo K.~Y.,  Fan X.,
    Strauss M.~A.,  2004, ApJ, 615, L17

\bibitem[\protect\citeauthoryear{{Weijmans}, {Krajnovi{\'c}}, {van de Ven},
  {Oosterloo}, {Morganti} \& {de Zeeuw}}{{Weijmans}
  et~al.}{2008}]{Weijmans:2008p3286}
{Weijmans} A.,  {Krajnovi{\'c}} D.,  {van de Ven} G.,  {Oosterloo} T.~A.,
  {Morganti} R.,    {de Zeeuw} P.~T.,  2008, MNRAS, 383, 1343

\bibitem[\protect\citeauthoryear{Williams, Bureau \& Cappellari}{Williams
  et~al.}{2009}]{Williams:2009p2681}
Williams M.~J.,  Bureau M.,    Cappellari M.,  2009, MNRAS, 400, 1665

\bibitem[\protect\citeauthoryear{Williams, Bureau \& Cappellari}{Williams
  et~al.}{2010}]{Williams2010}
Williams M.~J.,  Bureau M.,    Cappellari M.,  2010, MNRAS, submitted

\bibitem[\protect\citeauthoryear{Wrobel \& Kenney}{Wrobel \&
  Kenney}{1992}]{Wrobel:1992p982}
Wrobel J.~M.,  Kenney J. D.~P.,  1992, ApJ, 399, 94

\bibitem[\protect\citeauthoryear{Young}{Young}{2002}]{Young:2002p943}
Young L.~M.,  2002, AJ, 124, 788

\bibitem[\protect\citeauthoryear{Young, Bureau \& Cappellari}{Young
  et~al.}{2008}]{Young:2008p788}
Young L.~M.,  Bureau M.,    Cappellari M.,  2008, ApJ, 676, 317

\bibitem[\protect\citeauthoryear{Young, Bureau \& et al}{Young
  et~al.}{2011}]{Young2010}
Young L.~M.,  Bureau M.,    et al 2011, MNRAS (Paper IV)

\bibitem[\protect\citeauthoryear{Zwaan, van~der Hulst, de Blok \&
  McGaugh}{Zwaan et~al.}{1995}]{Zwaan:1995p2450}
Zwaan M.~A.,  van~der Hulst J.~M.,  de Blok W. J.~G.,    McGaugh S.~S.,  1995,
  MNRAS, 273, L35

\end{thebibliography}

\label{lastpage}

\end{document}